\documentclass[12pt,a4paper]{article}
  \usepackage{amsmath,amsthm,amssymb,hyperref}
  \allowdisplaybreaks
  \newtheorem{lemma}{Lemma}
  \newtheorem{cor}[lemma]{Corollary}
  
  \newtheorem{thm}[lemma]{Theorem}
  \newtheorem{rem}[lemma]{Remark}
  \newtheorem{obs}[lemma]{Observation}
  \newtheorem{main}{Theorem}
  
  \usepackage{verbatim}
  
  \usepackage[normalem]{ulem}
  
  \usepackage{listings}
  \usepackage{xcolor}  
  \definecolor{dkgreen}{rgb}{0.0,0.6,0.0}
  \definecolor{dkblue}{rgb}{0,0.1,0.7}
  \definecolor{dkred}{rgb}{0.5,0,0.3}
  \definecolor{gray}{rgb}{0.5,0.5,0.5}
  \definecolor{mauve}{rgb}{0.58,0,0.82}

  \definecolor{move_right}{rgb}{1.0,0.0,0.0}
  \definecolor{move_left}{rgb}{0.6,0.8,1.0}
  \definecolor{cycle}{rgb}{0.0,0.0,0.0}
  \definecolor{done}{rgb}{0.0,0.7,0.0}

  \usepackage[a4paper]{geometry}

  \newcommand{\pref}[1]{(\ref{#1})}

  \usepackage{tikz}
  \usetikzlibrary{datavisualization,plotmarks}
  
  \newcommand{\PairedDelimiters}[4]{\begingroup
    \def\empty{}\def\dummy{#1}\relax
    \ifx\empty\dummy
      \left#2#4\right#3\relax
    \else
      \dummy#2#4\dummy#3\relax
    \fi
    \endgroup
  }
  \newcommand{\Parens}[2][]{\PairedDelimiters{#1}{(}{)}{#2}}
  \newcommand{\FloorOf}[2][]{\PairedDelimiters{#1}{\lfloor}{\rfloor}{#2}}
  \newcommand{\EvalAt}[2][]{\negthinspace \PairedDelimiters{#1}{(}{)}{#2}}
  \newcommand{\Adjoin}[2][]{\negthinspace \PairedDelimiters{#1}{[}{]}{#2}}
  \newcommand{\AbsValueOf}[2][]{\PairedDelimiters{#1}{|}{|}{#2}}
  \newcommand{\Parentheses}[2][]{\Parens[#1]{#2}}
  \newcommand{\GaussOf}[2][]{\PairedDelimiters{#1}{\{}{\}}{#2}}
  
  \newcommand{\fibindex}{\kappa}

  \newcommand{\notion}{}
  \newcommand{\Cost}[1][]{m_{#1}}
  \newcommand{\CostOf}[2][]{\Cost[#1]\EvalAt{#2}}
  \newcommand{\EssentialCost}[1][]{\psi_{#1}}
  \newcommand{\EssentialCostOf}[2][]{\EssentialCost[#1]\EvalAt{#2}}
  \newcommand{\SimpleCost}[1][]{\mu_{#1}}
  \newcommand{\SimpleCostOf}[2][]{\SimpleCost[#1]\EvalAt{#2}}
  \newcommand{\RelCost}[1][]{f_{#1}}
  \newcommand{\RelCostOf}[2][]{\RelCost[#1]\EvalAt{#2}}
  \newcommand{\TheVariable}{x}
  \newcommand{\RndVariable}{X}
  \newcommand{\ExpValOf}[1]{\mathbb{E}\Adjoin{#1}}
  \newcommand{\TheOrder}{m}
  \newcommand{\RelLength}{x}
  \newcommand{\AltLength}{y}
  \newcommand{\Size}[1][]{n_{#1}}
  \newcommand{\Buffer}{b}
  \newcommand{\Order}{m}
  \newcommand{\Shift}[1][]{k_{#1}}
  \newcommand{\Blocks}{q}
  \newcommand{\EntryAt}[1]{a_{#1}}
  \newcommand{\TheTotal}{\nu}
  \newcommand{\TheLength}{\kappa}
  \newcommand{\BufLength}{\beta}
  \newcommand{\MovesOf}[2]{m(#1,#2)}
  \newcommand{\RemainderSumOf}[2]{\overline{m}(#1,#2)}
  \newcommand{\AvgCostOf}[1]{\rho(#1)}
  \newcommand{\TheCostConst}{D}
  \newcommand{\TheZetaConst}{C}
  \newcommand{\BigO}[1]{\operatorname{O}(#1)}
  
  \newcommand{\Fib}[1]{F_{#1}}
  \newcommand{\diff}{\operatorname{d}}
  \newcommand{\gcdOf}[1]{\gcd(#1)}
  \newcommand{\inm}[2]{(#1 \operatorname{mod} #2)}
  \newcommand{\ExpansionOf}[1]{[#1]}
  \newcommand{\Quotient}[1][]{q_{#1}}
  \newcommand{\Remainder}[1][]{r_{#1}}
  \newcommand{\PrevRem}[1][]{b_{#1}}
  \newcommand{\TheIndex}{j}
  \newcommand{\LastIndex}{t}
  \newcommand{\MidIndex}{s}
  \newcommand{\Entry}[1][]{c_{#1}}
  \newcommand{\LeftA}{x}
  \newcommand{\LeftB}{x'}
  \newcommand{\RightA}{y}
  \newcommand{\RightB}{y'}
  \newcommand{\SetOf}[2][]{\{#2\,\mid\,#1\}}
  \newcommand{\Sum}[2][]{\sum_{#1}#2}
  \newcommand{\InvQuantityOf}[1]{R(#1)}
  \newcommand{\TheQuantity}{G}
  \newcommand{\TheQuantityOf}[2][]{\TheQuantity_{#1}\EvalAt{#2}}
  \newcommand{\BulkQuantity}{\TheQuantity^{<}_{1}}
  \newcommand{\SmallQuantity}{\TheQuantity^{\geq}_{1}}
  \newcommand{\BulkQuantityOf}[1]{\BulkQuantity\Parens{#1}}
  \newcommand{\SmallQuantityOf}[1]{\SmallQuantity\Parens{#1}}  
  \newcommand{\divides}{|}
  \newcommand{\TheDivisor}{d}
  \newcommand{\MoebiusOf}[1]{\mu(#1)}
  \newcommand{\TheConst}{C}
  \newcommand{\AltDivisor}{f}
  \newcommand{\TheEpsilon}{\varepsilon}
  \newcommand{\congruent}{\equiv}
  \newcommand{\MinOf}[1]{\operatorname{min}\EvalAt{#1}}
  \newcommand{\TheCongruenceClass}{a}
  \newcommand{\ConstantTerm}{A}
  \newcommand{\LinearTerm}{B}
  \newcommand{\UpperBound}{Y}
  \newcommand{\UpperBoundOf}[1]{\UpperBound(#1)}
  \newcommand{\TheChar}{{\mathrm{e}_x}}
  \newcommand{\TheCharOf}[1]{\TheChar(#1)}
  \newcommand{\ModRing}{\ZZZ/\LeftA\ZZZ}
  \newcommand{\ModRingElement}{b}
  \newcommand{\AbsRingElement}{\AbsValueOf{\ModRingElement}}
  \newcommand{\mapcolon}{:}
  \newcommand{\ZZZ}{\mathbb{Z}}
  \newcommand{\CCC}{\mathbb{C}}
  \newcommand{\ExpOf}[1]{\mathrm{e}^{#1}}
  \newcommand{\SqrtOf}[1]{\sqrt{#1}}
  \newcommand{\AltIndex}{k}
  \newcommand{\TheAngle}{\phi}
  \newcommand{\ImUnit}{\mathrm{i}}
  \newcommand{\BotBound}{k}
  \newcommand{\TopBound}{l}
  \newcommand{\LogOf}[1]{\operatorname{ln}(#1)}
  
  \newcommand{\ZetaOf}[1]{\zeta\Parens{#1}}

  \newcommand{\NatNumbers}{\mathbb{N}}
  \newcommand{\NatExponent}{r}
  \newcommand{\InductionBound}{B}
  \newcommand{\Inside}[1][]{S^{#1}}
  \newcommand{\InsideOf}[2][]{\Inside[#1]\EvalAt{#2}}
  \newcommand{\Outside}{T}
  \newcommand{\OutsideOf}[1]{\Outside\EvalAt{#1}}
  \newcommand{\UpperBd}{N}
  \newcommand{\TheNatNumber}{M}
  \newcommand{\TheInterval}{I}
  
\begin{document}
  \title{The cost of cyclic permutations and remainder sums in the Euclidean algorithm}

  \author{Valentin Blomer \and Kai-Uwe Bux\thanks{Both authors thank the Deutsche
      Forschungsgemeinschaft (DFG, German Research Foundation) for support
      via the CRC TRR 358 – Project-ID 491392403.}
  }
  
  \date{January 2, 2026}

  \maketitle

  \begin{abstract}
    We discuss a modification to the Gries--Mills block swapping
    scheme for in-place rotation with average costs of 1.85 moves per
    element and worst case performance still at 3 moves per element.
    Analysis of the average case relies on the asymptotic behavior of the
    sum of remainders in the Euclidean algorithm.
  \end{abstract}

  \section{Introduction}\label{sec.intro}
  Rotating an array of length $\Size$ by $\Shift$ places performs a
  cyclic permutation of the entries, like cutting a deck of $\Size$
  cards lifting $\Shift$ of them and putting them back underneath. As
  any permutation, it can be written as a product of cycles; and the
  cycle decomposition of the rotation has $\gcd(\Size,\Shift)$ cycles,
  each of length $\frac{\Size}{\gcd(\Size,\Shift)}$. Recall that a
  cyclic permutation of order $\Order$ can be performed using $\Order+1$
  moves (using just one additional cell of memory). Hence performing a
  rotation this way takes $\Size+\gcd(\Size,\Shift)$ moves. This scheme
  for rotation is known as the Dolphin
  algorithm~\cite[Section~4]{GriesMills81}. Shene~\cite{Shene97} has
  shown that this number is optimal: any rotation algorithm will make at
  least that many moves.

  Despite being optimal, the Dolphin algorithm is slow on modern
  machines especially for large values of $\Size$. The algorithm
  accesses memory very much out of order; and for very large arrays, it
  approaches the point where every memory access is a cache miss. This
  motivates the search for rotation algorithms with better memory
  locality. Two such schemes have been discussed by
  Gries--Mills~\cite{GriesMills81}, predating the need of memory
  locallity:
  one may implement a rotation as a
  composition of three reversals, or one can employ a recursive strategy
  swapping segments of identical lengths at each step. The reversal
  scheme uses $\Size-O(1)$ swaps and the recursive block swapping
  strategy needs $\Size-\gcd(\Size,\Shift)$ swaps. Thus, both algorithms
  need on the order of $3\Size$ moves, in the latter case at least for
  ``typical'' $k$.

  Van~den~Hoven \cite{VanDenHoven21} has observed that some of the
  transpositions during a triple reversal can be rearranged into
  products of three or two, so that the rotation becomes a product of
  4-cycles, 3-cycles and transpositions. As three transpositions use
  nine moves, whereas a 4-cycles uses five moves, replacing three
  transpositions by a 4-cycle is a significant reduction in the number
  of moves. The resulting \notion{trinity rotation} uses essentially
  $2\Size$ moves, and performs significantly better than triple reverse.

  We introduce the block cycle scheme, an alternative to trinity
  rotation:
  \begin{main}
    Given an array of length $\Size$, 
    the block cycle rotation algorithm uses $3\Size$ moves in the
    worst case. For the average number of moves $\AvgCostOf{\Size}$,
    where the length $\Shift$ of the left segment is uniformly
    distributed between $0$ and $\Size$, we have
    \[
      \AvgCostOf{\Size} =
      \TheCostConst\Size + \BigO{ \Size^{\frac12 + \TheEpsilon } }
    \]
    for each $\TheEpsilon > 0$. The constant $\TheCostConst$ is
    approximately $1.85$.
  \end{main}
  The algorithm has good memory locality, uses sequential memory
  access patterns, and performs competitively in experiments.

  We note that triple reversal and trinity rotation will temporarily
  reverse (locally) the ordering of the items in the array. The block
  swapping and block cycling scheme, on the other hand, preserve the
  order of elements inside blocks. As a consequence, the algorithms show
  a clear distinction with regard to the dependency of the run time on
  the type of the underlying data. See section~\ref{sec.benchmarks} for
  more information. In particular, the block cycle algorithm seems
  particularly well suited for rotations in long words over small
  alphabets.

  We describe the block cycle scheme in section~\ref{sec.algorithm}.
  In section~\ref{sec.linear-term}, we show that $\AvgCostOf{\Size}$
  grows asymptotically linearly with $\Size$; and in
  section~\ref{sec.restglied}, we bound the error term.  This requires
  techniques from analytic number theory, in particular bounding certain
  exponential sums. The constant $D$ turns out to be the integral of
  a function that has certain self-similar features, but is
  Riemann-integrable. We compute the constant as an infinite series that
  can be expressed in term of multiple zeta values and estimated
  numerically.

  \paragraph{Acknowledgments}
  We thank Markus~Nebel and Jens~Stoye for helpful conversations and
  references. We thank Markus~Nebel and Pascal~Schweizer for feedback on
  an earlier draft of this paper.

  \section{The block cycle algorithm}\label{sec.algorithm}
  The starting point for the block cycle algorithm is the following
  block swapping scheme that slightly modifies the scheme of Gries and
  Mills: They swap the smaller segment, say of length $\Shift$, with its
  mirror image on the other end thereby moving all its elements into
  their final position and leaving a rotation problem of complementary
  size $\Size-\Shift$. In our variant, the smaller segment is swapped
  with the \emph{adjacent} block of the same length. This way, also
  $\Shift$ elements reach their final destination at the expense of
  $\Shift$ swaps.  The difference is that in the original Gries--Mills
  scheme, the \emph{elements} in the smaller segment are moved to their
  final destinations, whereas in our variation, the \emph{positions} in
  the smaller segment are filled with the elements that belong there. In
  either case, the remaining problem is a rotation in an array of length
  $\Size-\Shift$. It can be solved via tail recursion, where the
  termination condition is $\Shift=0$ or $\Shift=\Size$.

  This variation has exactly the same swap count as the original block
  swapping scheme. However, it lends itself more readily to combining
  swaps onto longer cycles. Specifically, we perform a rotation of the
  segments
  \[
    \EntryAt{1},\ldots,\EntryAt{\Shift}\mid
    \EntryAt{\Shift+1},\ldots,\EntryAt{2\Shift}\mid
    \EntryAt{2\Shift+1},\ldots,\EntryAt{3\Shift}\mid
    \ldots
    \mid \EntryAt{(\Blocks-1)\Shift+1},\ldots,\EntryAt{\Blocks\Shift}
  \]
  into the new order
  \[
    \EntryAt{\Shift+1},\ldots,\EntryAt{2\Shift}\mid
    \EntryAt{2\Shift+1},\ldots,\EntryAt{3\Shift}\mid
    \ldots
    \mid \EntryAt{(\Blocks-1)\Shift+1},\ldots,\EntryAt{\Blocks\Shift}
    \mid \EntryAt{1},\ldots,\EntryAt{\Shift}
  \]  
  where $\Blocks=\FloorOf{\frac{\Size}{\Shift}}$.  Note that we
  have moved $(\Blocks-1)\Shift$ elements into their final position
  using $(\Blocks+1)\Shift$ moves; and we have thus reduced the problem
  to another rotation within the segment
  $(\EntryAt{(\Blocks-1)\Shift+1},\ldots,\EntryAt{\Size})$, namely
  swapping
  $(\EntryAt{(\Blocks-1)\Shift+1},\ldots,\EntryAt{\Blocks\Shift})$ with
  $(\EntryAt{\Blocks\Shift+1},\ldots,\EntryAt{\Size})$. We can use tail
  recursion and solve the smaller remaining problem.
  
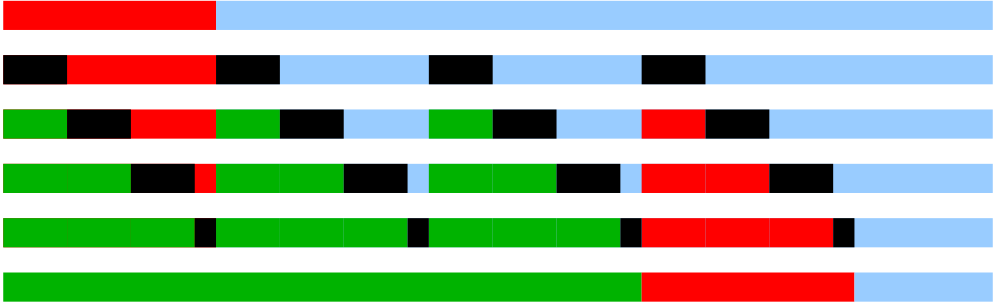
\begin{figure}[t]
    \begin{center}
      \begin{tikzpicture}[xscale=1.4,yscale=1.5]
        \draw[line width=11pt,color=move_right] (-1,0) -- (1,0);
        \draw[line width=11pt,color=move_left] (1,0) -- (8.3,0);
      \end{tikzpicture}
      
      \medskip

      \begin{tikzpicture}[xscale=1.4,yscale=1.5]
        \draw[line width=11pt,color=move_right] (-1,0) -- (1,0);
        \draw[line width=11pt,color=move_left] (1,0) -- (8.3,0);
        \draw[line width=11pt,color=cycle] (-1,0) -- (-0.4,0);
        \draw[line width=11pt,color=cycle] (1,0) -- (1.6,0);
        \draw[line width=11pt,color=cycle] (3,0) -- (3.6,0);
        \draw[line width=11pt,color=cycle] (5,0) -- (5.6,0);
      \end{tikzpicture}

      \medskip

      \begin{tikzpicture}[xscale=1.4,yscale=1.5]
        \draw[line width=11pt,color=move_right] (-1,0) -- (1,0);
        \draw[line width=11pt,color=move_left] (1,0) -- (8.3,0);
        \draw[line width=11pt,color=done] (-1,0) -- (-0.4,0);
        \draw[line width=11pt,color=done] (1,0) -- (1.6,0);
        \draw[line width=11pt,color=done] (3,0) -- (3.6,0);
        \draw[line width=11pt,color=move_right] (5,0) -- (5.6,0);
        \draw[line width=11pt,color=cycle] (-0.4,0) -- (0.2,0);
        \draw[line width=11pt,color=cycle] (1.6,0) -- (2.2,0);
        \draw[line width=11pt,color=cycle] (3.6,0) -- (4.2,0);
        \draw[line width=11pt,color=cycle] (5.6,0) -- (6.2,0);
      \end{tikzpicture}

      \medskip      

      \begin{tikzpicture}[xscale=1.4,yscale=1.5]
        \draw[line width=11pt,color=move_right] (-1,0) -- (1,0);
        \draw[line width=11pt,color=move_left] (1,0) -- (8.3,0);
        \draw[line width=11pt,color=done] (-1,0) -- (-0.4,0);
        \draw[line width=11pt,color=done] (1,0) -- (1.6,0);
        \draw[line width=11pt,color=done] (3,0) -- (3.6,0);
        \draw[line width=11pt,color=move_right] (5,0) -- (5.6,0);
        \draw[line width=11pt,color=done] (-0.4,0) -- (0.2,0);
        \draw[line width=11pt,color=done] (1.6,0) -- (2.2,0);
        \draw[line width=11pt,color=done] (3.6,0) -- (4.2,0);
        \draw[line width=11pt,color=move_right] (5.6,0) -- (6.2,0);
        \draw[line width=11pt,color=cycle] (0.2,0) -- (0.8,0);
        \draw[line width=11pt,color=cycle] (2.2,0) -- (2.8,0);
        \draw[line width=11pt,color=cycle] (4.2,0) -- (4.8,0);
        \draw[line width=11pt,color=cycle] (6.2,0) -- (6.8,0);
      \end{tikzpicture}

      \medskip     

      \begin{tikzpicture}[xscale=1.4,yscale=1.5]
        \draw[line width=11pt,color=move_right] (-1,0) -- (1,0);
        \draw[line width=11pt,color=move_left] (1,0) -- (8.3,0);
        \draw[line width=11pt,color=done] (-1,0) -- (-0.4,0);
        \draw[line width=11pt,color=done] (1,0) -- (1.6,0);
        \draw[line width=11pt,color=done] (3,0) -- (3.6,0);
        \draw[line width=11pt,color=move_right] (5,0) -- (5.6,0);
        \draw[line width=11pt,color=done] (-0.4,0) -- (0.2,0);
        \draw[line width=11pt,color=done] (1.6,0) -- (2.2,0);
        \draw[line width=11pt,color=done] (3.6,0) -- (4.2,0);
        \draw[line width=11pt,color=move_right] (5.6,0) -- (6.2,0);
        \draw[line width=11pt,color=done] (0.2,0) -- (0.8,0);
        \draw[line width=11pt,color=done] (2.2,0) -- (2.8,0);
        \draw[line width=11pt,color=done] (4.2,0) -- (4.8,0);
        \draw[line width=11pt,color=move_right] (6.2,0) -- (6.8,0);
        \draw[line width=11pt,color=cycle] (0.8,0) -- (1,0);
        \draw[line width=11pt,color=cycle] (2.8,0) -- (3,0);
        \draw[line width=11pt,color=cycle] (4.8,0) -- (5,0);
        \draw[line width=11pt,color=cycle] (6.8,0) -- (7,0);
      \end{tikzpicture}

      \medskip

      \begin{tikzpicture}[xscale=1.4,yscale=1.5]
        \draw[line width=11pt,color=done] (-1,0) -- (5,0);
        \draw[line width=11pt,color=move_right] (5,0) -- (7,0);
        \draw[line width=11pt,color=move_left] (7,0) -- (8.3,0);
      \end{tikzpicture}
    \end{center}
    \caption{\label{algorithm}%
      Recursive step of the block cycle algorithm:
      Red and blue lines represent items to be moved to the right and to the
      left, respectively. Black segments are currently undergoing a cyclic
      block permutation. The length of such a segment must not exceed
      $\Buffer$. Green lines indicate elements that have reached their final
      destination.}
  \end{figure}

  For large $\Blocks$, the block cycle resembles the Dolphin algorithm. To
  improve memory access patterns, this block rotation is performed in
  batches of adjacent elements of up to $\Buffer$.
  See~Figure~\ref{algorithm} for an illustration of the process with
  $\Blocks=4$.
  
  The fixed size buffer space accommodating up to $\Buffer$ items can
  also be used to end the recursion early.  We exit the
  recursion as soon as the shorter segment (of length $\Shift$) fits
  into the buffer: then, we can move the shorter segment it in its entirety
  into the buffer; now, we move all remaining $\Size-\Shift$ elements
  into their final position; finally, we move the $\Shift$ elements from
  the buffer into their target positions. We have thus performed the
  final rotation in $\Size+\Shift$ moves.

  Instead of pseudocode, we shall provide a reference implementation
  as a generic in \verb-C++- in section~\ref{sec.benchmarks} that we
  benchmark in comparison to other rotation algorithms.
  \begin{rem}
    The idea of combining swaps of the Gries--Mills scheme into long
    cycles and to move elements in batches are also the main points of the
    blend rotation scheme proposed by
    Hashem--Li--Salah~\cite[Algorithm~3]{HashemLiSalah23}.  They, however,
    do not recurse, instead they just do the first step and then solve the
    remaining problem using any other rotation method (they suggest triple
    reversal).  Effectively, they propose to prefix a fall back rotation
    method with the step shown in Figure~\ref{algorithm}.

    Moreover, they do not use symmetry. Consequently, their algorithm
    immediately resorts to the fall back method when the left segment is
    longer than the right segment, which leaves a sizable
    performance gain on the table. It appears likely that the
    pseudocode does not accurately reflect what they actually have
    benchmarked. Another hint in this direction is the fact that the
    pseudocode for blend rotation does not correctly handle the case that
    the buffer size does not divide evenly into the size of the left
    segment.
  \end{rem}

  \section{Preliminary analysis of worst and average case}\label{sec.linear-term}
  As the block cycle scheme is just combining swaps from the block
  swapping scheme into longer cycles, it clearly does not use more moves
  than the Gries--Mills scheme. Thus, we already have a bound for the
  worst case.
  \begin{obs}
    The block cycle algorithm uses at most $3\Size - 3\gcd(\Size,\Shift)$ moves. \qed
  \end{obs}
  \begin{obs}\label{fib-worst}
    The worst case materializes for instance, when the left segment
    has length $8$ and the right segment has length $13$.  Then, the
    recursion goes like so: $8:13$, $8:5$, $3:5$, $3:2$, $1:2$,
    and hence takes $3\cdot(8+5+3+2) + 4 = 58$ steps.
    More generally, this happens (asymptotically) for Fibonacci
    numbers $\Shift = F_{\fibindex}$, $\Size = F_{\fibindex+2}$ with
    $\gcdOf{\Size,\Shift} = 1$, where the algorithm takes
    $3 \sum_{j=1}^{\fibindex} F_j - 2 = 3 \Size - 3\gcdOf{\Size,\Shift} - 2$ steps.
  \end{obs}

  Let $\CostOf{\Size,\Shift,\Buffer}$ denote the number of moves the
  block cycle method needs to rotate $\Shift$ steps to the left in an
  array of length $\Size$ using a buffer of size $\Buffer$.
  For $\Shift \leq \frac{\Size}{2}$, we find
  \begin{equation}\label{integral}
    \CostOf{\Size,\Shift,\Buffer}
    =
    \begin{cases}
      0 & \text{if\ } \Shift = 0, \\
      \Size+\Shift & \text{if\ } \Shift \leq \Buffer, \\
      \Parentheses{\FloorOf{ \frac{\Size}{\Shift} } + 1} \Shift
      +
      \CostOf{
      \Size - \Parentheses{\FloorOf{\frac{\Size}{\Shift}} - 1} \Shift,
      \Size - \FloorOf{\frac{\Size}{\Shift}} \Shift,
      \Buffer
      }
        & \text{otherwise}.
    \end{cases}
  \end{equation}
  By symmetry of left rotation and right rotation, we have
  \(
    \CostOf{\Size,\Shift,\Buffer} = \CostOf{\Size,\Size-\Shift,\Buffer}
  \)
  for $\Shift \geq \frac{\Size}{2}$. 
  
  It is more convenient to consider a slightly simplified version,
  whose definition extends to real valued variables, which we denote by
  greek letters as a visual clue. For $\TheTotal, \BufLength >0$ and
  $\TheLength \in [0,\frac{\TheTotal}{2}]$, we define
  \begin{equation}\label{continuous}
    \SimpleCostOf{\TheTotal,\TheLength,\BufLength} :=
    \begin{cases}
      \TheTotal + \TheLength & \text{if\ } \TheLength \leq \BufLength, \\
      (\FloorOf{ \frac{\TheTotal}{\TheLength} } + 1) \TheLength
      +
      \SimpleCostOf{
      \TheTotal
      - (\FloorOf{ \frac{\TheTotal}{\TheLength} } - 1)
      \TheLength,
      \TheTotal - \FloorOf{ \frac{\TheTotal}{\TheLength} } \TheLength, \BufLength }
                             & \text{otherwise}.
    \end{cases}
  \end{equation}
  We extend to $\TheLength \in [0,\TheTotal]$ by symmetry
  \(
    \SimpleCostOf{\TheTotal,\TheLength,\BufLength}=
    \SimpleCostOf{\TheTotal,\TheTotal-\TheLength,\BufLength}
    .
  \)
  
  For $\TheTotal>0$ and $\TheLength\in[0,\frac{\TheTotal}{2}]$, we put
  \[
    \TheLength'
    := \TheTotal - \FloorOf{\frac{\TheTotal}{\TheLength} } \TheLength
    =
    \TheLength\Parentheses{
      \frac{\TheTotal}{\TheLength} - \FloorOf{ \frac{\TheTotal}{\TheLength} }
    }
    =
    \TheLength\GaussOf{\frac{\TheTotal}{\TheLength}}
    \in[0, \TheLength]
  \]
  and
  \[
    \TheTotal'
    := \TheTotal -
    \Parentheses{ \FloorOf{ \frac{\TheTotal}{\TheLength} } - 1 } \TheLength
    =
    \TheLength + \TheLength'
    =
    \TheLength\Parentheses{1+\GaussOf{\frac{\TheTotal}{\TheLength}}}
  \]
  where we make use of the \notion{Gauss map}
  $\GaussOf{\TheVariable}=\TheVariable-\FloorOf{\TheVariable}$.
  Then, we find
  \begin{equation}\label{def-mu-nu}
    \SimpleCostOf{ \TheTotal, \TheLength, \BufLength }
    -
    \TheTotal
    =
    \begin{cases}
      \TheLength & \text{if\ } \TheLength \leq \BufLength, \\
      2\TheLength
      +
      \SimpleCostOf{\TheTotal',\TheLength',\BufLength}
      -
      \TheTotal'
                 & \text{otherwise}.
    \end{cases}
  \end{equation}

  \begin{obs}\label{geometric-decay}
    For $\TheTotal>0$ and $0<\TheLength\leq\frac{\TheTotal}{2}$, we have
    \[
      \frac{\TheTotal'}{\TheTotal}
      =
      \frac{\TheLength}{\TheTotal}
      \Parentheses{1+\GaussOf{{\displaystyle\frac{\TheTotal}{\TheLength}}}}
      \leq \frac23,
    \]
    which is obvious for $\TheLength\leq\frac{\TheTotal}{3}$
    and can be seen for $\frac{\TheTotal}{3}<\TheLength\leq\frac{\TheTotal}{2}$
    by writing
    \(
      \frac{\TheLength}{\TheTotal} = \frac{1}{2+\TheVariable}
    \)
    with $0\leq\TheVariable\leq1$,
    which leads to
    \(
      \frac{\TheLength}{\TheTotal}
      \Parentheses{1+\GaussOf{{\frac{\TheTotal}{\TheLength}}}}
      =
      \TheVariable\Parentheses{1+\GaussOf{\frac{1}{\TheVariable}}}
      =
      \frac{1+\TheVariable}{2+\TheVariable}\leq\frac23
    \).\qed
  \end{obs}
  Observation~\ref{geometric-decay} allows us to prove statements
  about $\SimpleCost$ and $\EssentialCost$ by induction on the smallest
  $\NatExponent\in\NatNumbers$ satisfying
  $\TheTotal\leq\frac{3^\NatExponent}{2^\NatExponent}\BufLength$.  So
  the base case is $\TheTotal\leq\BufLength$ and the induction step goes
  from $\TheTotal\leq\InductionBound$ to
  $\TheTotal\leq\frac32\InductionBound$.

  As a demonstration of the method, we show that
  the worst case estimate also holds in the continuous setting:
  \begin{obs}
    We have $\SimpleCostOf{\TheTotal,\TheLength,\BufLength}\leq 3\TheTotal$.
  \end{obs}
  \begin{proof}
    The claim is clear for $\TheTotal\leq \BufLength$, because then, we have
    $\SimpleCostOf{\TheTotal,\TheLength,\BufLength} = \TheTotal+\TheLength \leq 3\TheTotal$.

    For $\TheTotal>\BufLength$, we have
    $\SimpleCostOf{\TheTotal,\TheLength,\BufLength}=\TheTotal+\TheLength\leq
    3\TheTotal$ if $\TheLength\leq\BufLength$ and
    \[
      \SimpleCostOf{\TheTotal,\TheLength,\BufLength}
      = \FloorOf{\frac{\TheTotal}{\TheLength}+1} \TheLength +
      \SimpleCostOf{\TheTotal',\TheLength',\BufLength}
      = \frac{
        \FloorOf{\frac{\TheTotal}{\TheLength}+1}
      }{
        \FloorOf{\frac{\TheTotal}{\TheLength}-1}
      }
      \FloorOf{\frac{\TheTotal}{\TheLength}-1}\TheLength
      + \SimpleCostOf{\TheTotal',\TheLength',\BufLength}
    \]
    otherwise. By induction on $n$ in
    $\TheTotal\leq\frac{3^n}{2^n}\BufLength$, we may assume that
    \(
      \SimpleCostOf{\TheTotal',\TheLength',\BufLength} \leq 3\TheTotal'
    \).
    Since
    \(
      \frac{
        \FloorOf{\frac{\TheTotal}{\TheLength}+1}
      }{\FloorOf{\frac{\TheTotal}{\TheLength}-1}
      } \leq 3
    \), we find
    \[
      \SimpleCostOf{\TheTotal,\TheLength,\BufLength}
      \leq
      3 \FloorOf{\frac{\TheTotal}{\TheLength}-1}\TheLength + 3 \TheTotal'
      = 3 \TheTotal
    \]
    as $\TheTotal' + \FloorOf{\frac{\TheTotal}{\TheLength}-1}\TheLength = \TheTotal$.
  \end{proof}
  The same line of argument also proves the following claims.
  \begin{cor}
    \begin{enumerate}
      \item
        The function $\SimpleCostOf{\TheTotal,\TheLength,\BufLength}$
        decreases monotonically in $\BufLength$.
      \item
        For $\lambda>0$, we have
        \(
        \SimpleCostOf{\lambda\TheTotal,\lambda\TheLength,\lambda\BufLength}
        =\lambda \SimpleCostOf{\TheTotal,\TheLength,\BufLength}
        \).
      \item
        The function $\SimpleCost$ provides an upper bound for the number of moves:
        \[
        \CostOf{\Size,\Shift,\Buffer} \leq \SimpleCostOf{\Size,\Shift,\Buffer}.
        \]
    \end{enumerate}
  \end{cor}

  In particular, as $\BufLength$ tends to $0_+$, the value
  $\SimpleCostOf{\TheTotal,\TheLength,\BufLength}$ is monotonically increasing, yet bounded from
  above by $3\TheTotal$.  Consequently, we can pass to the limit
  \[
    \SimpleCostOf{\TheTotal,\TheLength} =
    \lim_{\BufLength\rightarrow 0_+} \SimpleCostOf{\TheTotal,\TheLength,\BufLength}
    .
  \]
  
  Obviously, we have $\SimpleCostOf{\TheTotal,\TheLength,\BufLength}
  \leq \SimpleCostOf{\TheTotal,\TheLength}$ and
  $\SimpleCostOf{\lambda\TheTotal,\lambda\TheLength}
  =\lambda \SimpleCostOf{\TheTotal,\TheLength}
  .
  $
  Hence, all the information is contained in the function
  \[
    \RelCostOf{\TheLength} := \SimpleCostOf{1,\TheLength},
  \]
  which inherits the symmetry
  $\RelCostOf{\TheLength} = \RelCostOf{1-\TheLength}$. 

\begin{figure}[t]
    \begin{center}
      \begin{tikzpicture}[xscale=21,yscale=2,domain=0:0.5]
        \draw[thick] plot[smooth] file{rotate_cost.data};
        \draw[step=0.1,very thin,color=gray] (0,0) grid (0.5,3);
        \draw[thin,color=black] (0,1) -- (0.5,1);
        \draw[thin,color=black] (0,2) -- (0.5,2);
        \draw[thin,color=black] (0,3) -- (0.5,3);
        \draw (0,1) node[left] {$1$};
        \draw (0,2) node[left] {$2$};
        \draw (0,3) node[left] {$3$};
        \draw (0.5,0) node[below] {$\frac12$};
        \draw[->] (-0.04,0) -- (0.54,0) node[right] {$\RelLength$};
        \draw[->] (0,-0.2) -- (0,3.2) node[above] {$\RelCost$};
      \end{tikzpicture}
    \end{center}
    \caption{\label{plot-f}
      The function $\RelCost$ over the range $0\leq\RelLength\leq
      \frac12$. The maximum value $3$ is attained at
      $\frac{3-\sqrt{5}}{2}=1-\phi$ where $\phi$ is the golden mean.
      Let $(\Fib{m})$ denote the
      Fibonacci sequence. As
      \(
        1-\phi
        = \lim_{m\rightarrow\infty} \frac{
          \Fib{m-1}
        }{
          \Fib{m}+\Fib{m-1}
        }
      \)
      this peak corresponds to the worst case behavior of the algorithm
      noted in Observation~\ref{fib-worst}.
    }
  \end{figure}
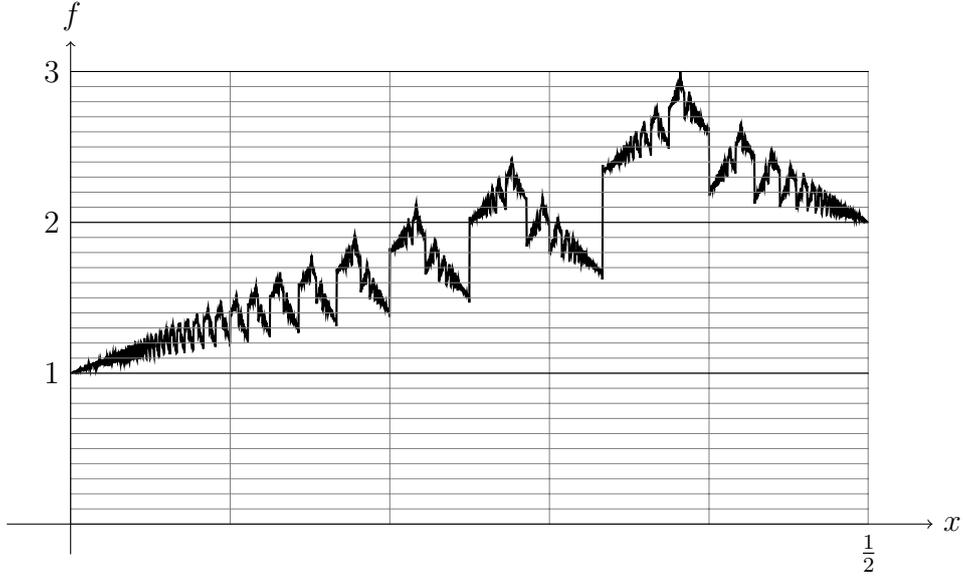

  We sketch the graph in Figure~\ref{plot-f}, which indicates that
  $\RelCost$ is somewhat self-similar.
  To derive self-similarities, note that for $0<\TheLength<\frac12$, we
  find, for instance, $\frac15<\frac{1-\TheLength}{4-3\TheLength}<\frac14$, whence
  \begin{align*}
    \RelCostOf{
    \frac{1-\TheLength}{4-3\TheLength}
    }
    &=
      5\frac{1-\TheLength}{4-3\TheLength}
      +
      \SimpleCostOf{
      \frac{1}{4-3\TheLength}
      ,
      \frac{\TheLength}{4-3\TheLength}
      }
      =
      \frac{
      5-5\TheLength
      +
      \RelCostOf{
      \TheLength
      }
      }{
      4-3\TheLength
      }
      .
  \end{align*}
  Solving for $\RelCostOf{\TheLength}$ yields the self-similarity
  \(
    \RelCostOf{\TheLength} =
    5\TheLength-5 + (4-3\TheLength)\RelCostOf{\frac{1-\TheLength}{4-3\TheLength}}
    .
  \)

  The same calculation for the range
  $\frac{1}{m+2}< \frac{1-\TheLength}{m+1-m\TheLength} < \frac{1}{m+1}$ yields
  the self-similarity
  \begin{equation}\label{reverse}
    \RelCostOf{\AltLength}
    = (m+2)\AltLength -(m+2) + (m+1-m\AltLength)
    \RelCostOf{\frac{1-\AltLength}{m + 1 - m\AltLength}}
  \end{equation}
  for $0<\TheLength<\frac12$.
  
  \begin{thm}\label{thm.integrable}
    The function $\RelCost$ is continuous at all irrational numbers.
    In particular, it has at most countably many places of discontinuity
    and is Riemann integrable.
  \end{thm}
  \begin{proof}
    The definition \eqref{def-mu-nu} suggests to define
    $\EssentialCostOf{\TheTotal,\TheLength,\BufLength}=
    \SimpleCostOf{\TheTotal,\TheLength,\BufLength} - \TheTotal$
    and correspondingly   
    \begin{equation}\label{defpsi}
      \EssentialCostOf{\TheTotal,\TheLength}
      :=
      \lim_{\BufLength\rightarrow 0_+}
      \EssentialCostOf{\TheTotal,\TheLength,\BufLength}
      =
      \SimpleCostOf{\TheTotal,\TheLength}-\TheTotal
      \quad\text{and}\quad
      \EssentialCostOf{\TheLength}:= \EssentialCostOf{1, \TheLength},
    \end{equation}
    and then to show that $\TheLength\mapsto\EssentialCostOf{\TheLength}$ is
    continuous at any irrational $\TheLength$.

    Note that the recurrence
    \begin{equation}\label{essential-recurrence}
      \EssentialCostOf{ \TheTotal, \TheLength, \BufLength }
      =
      \begin{cases}
        \TheLength & \text{if\ } \TheLength \leq \BufLength, \\
        2\TheLength
        +
        \EssentialCostOf{\TheTotal',\TheLength',\BufLength}
                   & \text{otherwise}
      \end{cases}
    \end{equation}
    describes a recursive algorithm for
    computing $\EssentialCostOf{\TheTotal,\TheLength,\BufLength}$
    with termination upon $\TheLength \leq \BufLength$. As $\BufLength$ tends to $0_+$,
    the recursion depth increases.  In the limit $\BufLength=0$, we find that the recursion only
    terminates when $\TheLength$ and $\TheTotal$ are commensurable.  Thus
    for rational $\frac{\TheLength}{\TheTotal}$, we can expand
    $\EssentialCostOf{ \TheTotal, \TheLength }$ into a finite sum.  If
    $\frac{\TheLength}{\TheTotal}$ starts out irrational, it will stay
    irrational during the recursion and the termination branch will never
    be taken. In this case, the infinite recursion expresses
    $\EssentialCostOf{ \TheTotal, \TheLength }$ as a convergent series.

    We use homogeneity to deduce
    \[
      \EssentialCostOf{\TheLength}
      =
      \begin{cases}
        0 & \TheLength = 0, \\
        2\TheLength +
        \TheLength
        \Parentheses{1+\GaussOf{\frac{1}{\TheLength}}}
        \EssentialCostOf{
        \TheLength\GaussOf{\frac{1}{\TheLength}}}
          & \text{otherwise}.
      \end{cases}
    \]
    Using the definitions
    \begin{align*}
      \OutsideOf{\TheVariable}
      := \TheVariable
      \Parentheses{1+\GaussOf{\frac{1}{\TheVariable}}}, \quad 
      \InsideOf{\TheVariable}
      := \frac{
      \GaussOf{\frac{1}{\TheVariable}}
      }{
      1+\GaussOf{\frac{1}{\TheVariable}}
      }
    \end{align*}
    where we formally put $\InsideOf{0} = \OutsideOf{0} = 0$, we can
    thus remove the case distinction and write
    \[
      \EssentialCostOf{\TheVariable} =
      2 \TheVariable
      +
      \OutsideOf{\TheVariable}\EssentialCostOf{\InsideOf{\TheVariable}}
      .
    \]
    Unraveling the implicit recursion yields
    \begin{align*}
      \EssentialCostOf{\TheVariable}
      & =
        2 \TheVariable +
        2\sum_{\TheIndex=1}^{\infty}
        \OutsideOf{\TheVariable}
        \OutsideOf{\InsideOf{\TheVariable}}
        \cdots
        \OutsideOf{\InsideOf[\TheIndex-1]{\TheVariable}}
        \InsideOf[\TheIndex]{\TheVariable}
        .
    \end{align*}

    Note that $\OutsideOf{\InsideOf[\TheIndex]{\TheVariable}}\leq \frac23$
    and $\InsideOf[\TheIndex+1]{\TheVariable}\leq\frac12$.
    It follows that for every $\TheEpsilon>0$ there
    is $\UpperBd\in\NatNumbers$ such that
    \[
      2\sum_{\TheIndex>\UpperBd}
      \OutsideOf{\TheVariable}\OutsideOf{\InsideOf{\TheVariable}}
      \cdots\OutsideOf{\InsideOf[\TheIndex-1]{\TheVariable}}
      \InsideOf[\TheIndex]{\TheVariable}
      \leq
      \sum_{\TheIndex>\UpperBd}\Parentheses{\frac23}^{\TheIndex}
      \leq
      \TheEpsilon
    \]
    uniformly in $\TheVariable$. I.e., the
    series defining $\EssentialCostOf{\TheVariable}$ converges
    uniformly to $\EssentialCost$.

    Now, let $\TheVariable\in[0,\frac12]$ be irrational.  The function
    $\Inside$ and $\Outside$ are continuous on each open interval
    $(\frac{1}{\TheNatNumber+1},\frac{1}{\TheNatNumber})$. By considering
    $\UpperBd$-th generation iterated preimages, we find that there is an
    open interval $\TheInterval$ containing $\TheVariable$ on which the
    partial sum
    \[
      2 \TheVariable +
      2\sum_{\TheIndex=1}^{\UpperBd}
      \OutsideOf{\TheVariable}
      \OutsideOf{\InsideOf{\TheVariable}}
      \cdots
      \OutsideOf{\InsideOf[\TheIndex-1]{\TheVariable}}
      \InsideOf[\TheIndex]{\TheVariable}
    \]
    is continuous.

    Hence, $\EssentialCost$ is the uniform limit of functions continuous
    at $\TheVariable$. The claim follows.
  \end{proof}

  \begin{cor}
    Assume that $\RndVariable$ is a real random variable uniformly
    distributed in $[0,\frac12]$.
    Then the $\TheOrder$-th moment of the distribution of
    $\RelCostOf{\RndVariable}=\EssentialCostOf{\RndVariable}+1$
    is given as
    \[
      \ExpValOf{ \RelCostOf{\RndVariable}^\TheOrder }
      =
      \frac{\int_0^{\frac12} \RelCostOf{\TheVariable}^{\TheOrder} \diff \TheVariable}{\frac12}
      .
    \]
    In particular, moments of all orders exist.\qed
  \end{cor}

  We are now ready to show that the average number of moves for the block cycle
  algorithm has a well defined asymptotic behavior. We ignore the
  savings associated with using a large buffer: asymptotically any buffer of
  fixed size is insignificant. So let
  \[
    \MovesOf{\Size}{\Shift} = \CostOf{\Size,\Shift,1}
  \]
  denote the number of moves used by the block cycle algorithm for rotating
  an array of length $n$ by $k$ steps using a buffer of size~$1$.
  The average move count for rotations in an array of length $\Size$ is
  \[
    \AvgCostOf{\Size}
    :=
    \frac{1}{\Size}\sum_{0\leq \Shift < \Size}
    \MovesOf{\Size}{\Shift}
    ,
  \]
  where we assume the uniform distribution for $\Shift$.

  \begin{thm}\label{thm.limit}
    The average move count $\AvgCostOf{\Size}$ grows linearly with $\Size$.
    More precisely,
    \[
      \lim_{\Size\rightarrow\infty}
      \frac{\AvgCostOf{\Size}}{\Size}
      =
      \ExpValOf{ \RelCostOf{\RndVariable} }
      =
      2\int_0^{\frac12} \RelCostOf{\TheVariable} \diff \TheVariable
    \]
    where  $\RndVariable$ is a real random variable uniformly
    distributed in $[0,\frac12]$.
  \end{thm}
  Before we give the proof, we collect some preliminary observations.
  An evaluation of the constant will be given in the next section. 

  We distinguish two types of moves: the moves into and out of the
  buffer region (type~A) and the moves within the array (type~B).
  \begin{obs}\label{obs.typeA}
    For $\Shift\leq\frac{\Size}{2}$, let $(\Shift[\TheIndex],\Size[\TheIndex])$ be
    the parameters for the (sub)problem in the $\TheIndex$-th iteration of the block
    cycle algorithm. 
    The number of type~A moves is clearly
    \(
      2(\Shift[1] + \Shift[2] + \Shift[3] + \cdots)
    \).
    We have the following recursion:
    \begin{align*}
      & k_1 = k, \quad &n_1 = n, \\
      & k_{i+1} = \inm{n_i}{k_i}, \quad & n_{i+1} = k_i + k_{i+1}.
    \end{align*}
    This realizes the recursion~\pref{essential-recurrence} for the function $\EssentialCost$ with
    $\BufLength=0$.
    Consequently, the number of type~A moves equals
    \(
      \EssentialCostOf{\Size,\Shift}
      =
      \Size\EssentialCostOf{\frac{\Shift}{\Size}}
    \), using the notation \eqref{defpsi}. \qed    
  \end{obs}
  For future reference, we record that the sequence of segment sizes $(\Shift[\TheIndex])$
  is just the sequence of remainders encountered in a run of the Euclidean algorithm.
  \begin{lemma}\label{lemma.moves}
    Let $\Shift \leq \frac{\Size}{2}$. As above, let $(\Shift[\TheIndex],\Size[\TheIndex])$
    be the the parameters for the (sub)problem in the $\TheIndex$-th iteration of the block
    cycle algorithm. Consider the sequence $(\Remainder[\TheIndex])$ defined by
    \begin{align*}
      & \Remainder[1] = k, \quad  
      & \PrevRem[1] = n, \\
      & \Remainder[\TheIndex+1] = \inm{\PrevRem[\TheIndex]}{\Remainder[\TheIndex]} , \quad 
      & \PrevRem[\TheIndex+1]  = \Remainder[\TheIndex]
    \end{align*}
    recursively defined by the Euclidean algorithm
    that terminates when $\Remainder[\TheIndex+1] = 0$.

    Then $\Shift[\TheIndex]=\Remainder[\TheIndex]$ for each
    $\TheIndex$. In particular, 
    \begin{enumerate}
      \item
        The block cycle algorithm terminates on a
        subproblem with parameters $(0,\gcd(\Size,\Shift))$.
      \item
        Since $\Shift[1]+\Shift[2]+\cdots = \Remainder[1]+\Remainder[2]+\cdots$, we have 
        \[
        \Size\EssentialCostOf{\frac{\Shift}{\Size}}
        =
        \EssentialCostOf{\Size,\Shift}
        =
        2\Parentheses{\Remainder[1]+\Remainder[2]+\cdots}
        .
        \]
    \end{enumerate}
  \end{lemma}
  \begin{proof}
    A straightforward induction along the recursions for
    $(\Remainder[\TheIndex],\PrevRem[\TheIndex])$ and
    $(\Shift[\TheIndex],\Size[\TheIndex])$ shows \(
      \Remainder[\TheIndex] = \Shift[\TheIndex]$ and $\PrevRem[\TheIndex]
      \cong \Size[\TheIndex] \mod \Remainder[\TheIndex] \).
  \end{proof}

  \begin{obs}
    With each move of type~B, exactly one element reaches its final
    position.  Thus, there are at most $\Size$ moves of this type. In
    fact, each element has to undergo exactly one move of type~B except at
    the end of the recursion, when we hit a subproblem of rotating an
    array of size $\gcd(\Shift,\Size)$ by $0$. Hence, we make use of
    exactly $\Size - \gcd(\Shift,\Size)$ moves of type~B.\qed
  \end{obs}
  
  \begin{proof}[Proof of Theorem~\ref{thm.limit}]
    The move count $\MovesOf{\Size}{\Shift}$ is the sum of the type~A
    count and the type~B count. Hence, for
    $\Shift\leq\frac{\Size}{2}$, we have 
    \begin{equation}\label{relation}
      \MovesOf{\Size}{\Shift}
      = \Size - \gcd(\Shift,\Size) + \Size\EssentialCostOf{\frac{\Shift}{\Size}}
      = \Size\RelCostOf{\frac{\Shift}{\Size}} - \gcd(\Shift,\Size)
      .
    \end{equation}
    As $\gcd(\Shift,\Size) = \gcd(\Size-\Shift,\Size)$ and
    $\MovesOf{\Shift}{\Size}=\MovesOf{\Size-\Shift}{\Size}$, the formula stays valid for
    all $\Shift<\Size$; and we have
    \[
      2
      \int_0^{\frac12} \RelCostOf{\TheVariable} \diff \TheVariable
      =
      \int_0^1 \RelCostOf{\TheVariable} \diff \TheVariable
      =
      \lim_{\Size\rightarrow\infty} \frac{1}{\Size}
      \sum_{0\leq\Shift<\Size} \RelCostOf{\frac{\Shift}{\Size}}
      .
    \]
    
    Thus, we can compute
    \begin{align*}
      \lim_{\Size\rightarrow\infty} \frac{\AvgCostOf{\Size}}{\Size}
      & =
        \lim_{\Size\rightarrow\infty}
        \frac{1}{\Size^2}\sum_{0\leq\Shift<\Size}\MovesOf{\Shift}{\Size}
      \\
      & =
        \lim_{\Size\rightarrow\infty}
        \frac{1}{\Size^2}
        \sum_{0\leq\Shift<\Size}\Size\RelCostOf{\frac{\Shift}{\Size}}
        -
        \lim_{\Size\rightarrow\infty} \frac{1}{\Size^2}
        \sum_{0\leq\Shift<\Size}\gcd(\Shift,\Size)      
      \\
      & =
        2 \int_0^{\frac12} \RelCostOf{\TheVariable} \diff\TheVariable
        .
    \end{align*}
    This completes the proof.
  \end{proof}

  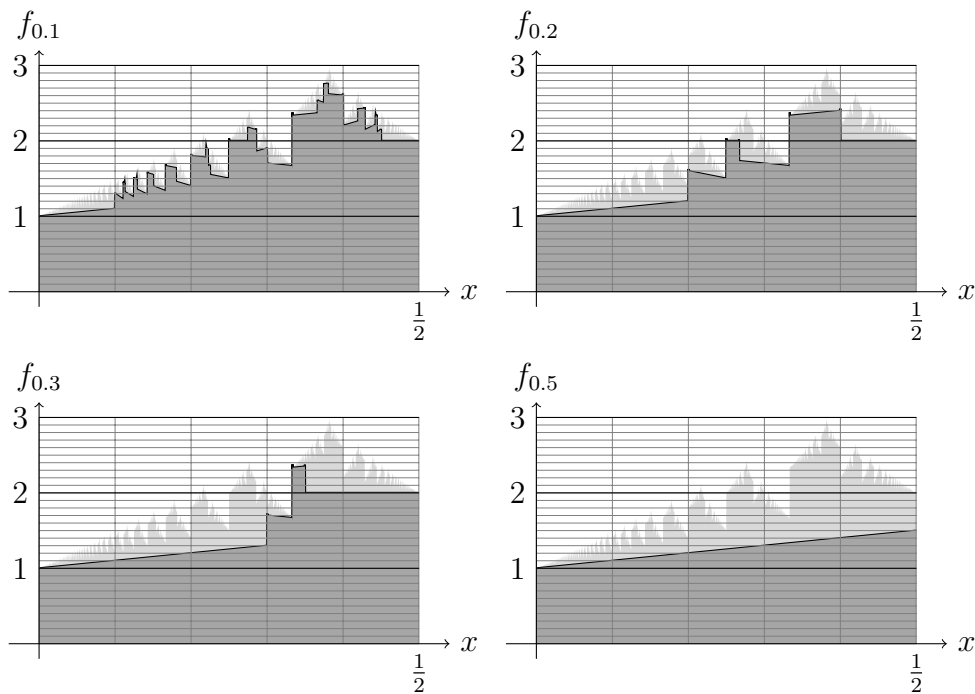
\begin{figure}[t]
    \begin{center}
      \begin{tikzpicture}[xscale=10,yscale=1,domain=0:0.5]
        \fill[color=gray!30] (0,0) -- plot[smooth] file{rotate_cost.data} -- (0.5,0) -- cycle;
        \draw[thick] plot[smooth] file{rotate_cost_10.data};
        \fill[color=gray!70] (0,0) -- plot[smooth] file{rotate_cost_10.data} -- (0.5,0) -- cycle;
        \draw[step=0.1,very thin,color=gray] (0,0) grid (0.5,3);
        \draw[thin,color=black] (0,1) -- (0.5,1);
        \draw[thin,color=black] (0,2) -- (0.5,2);
        \draw[thin,color=black] (0,3) -- (0.5,3);
        \draw (0,1) node[left] {$1$};
        \draw (0,2) node[left] {$2$};
        \draw (0,3) node[left] {$3$};
        \draw (0.5,0) node[below] {$\frac12$};
        \draw[->] (-0.04,0) -- (0.54,0) node[right] {$\RelLength$};
        \draw[->] (0,-0.2) -- (0,3.2) node[above] {$\RelCost[0.1]$};
      \end{tikzpicture}
      \begin{tikzpicture}[xscale=10,yscale=1,domain=0:0.5]
        \fill[color=gray!30] (0,0) -- plot[smooth] file{rotate_cost.data} -- (0.5,0) -- cycle;
        \draw[thick] plot[smooth] file{rotate_cost_20.data};
        \fill[color=gray!70] (0,0) -- plot[smooth] file{rotate_cost_20.data} -- (0.5,0) -- cycle;
        \draw[step=0.1,very thin,color=gray] (0,0) grid (0.5,3);
        \draw[thin,color=black] (0,1) -- (0.5,1);
        \draw[thin,color=black] (0,2) -- (0.5,2);
        \draw[thin,color=black] (0,3) -- (0.5,3);
        \draw (0,1) node[left] {$1$};
        \draw (0,2) node[left] {$2$};
        \draw (0,3) node[left] {$3$};
        \draw (0.5,0) node[below] {$\frac12$};
        \draw[->] (-0.04,0) -- (0.54,0) node[right] {$\RelLength$};
        \draw[->] (0,-0.2) -- (0,3.2) node[above] {$\RelCost[0.2]$};
      \end{tikzpicture}
      
      \begin{tikzpicture}[xscale=10,yscale=1,domain=0:0.5]
        \fill[color=gray!30] (0,0) -- plot[smooth] file{rotate_cost.data} -- (0.5,0) -- cycle;
        \draw[thick] plot[smooth] file{rotate_cost_30.data};
        \fill[color=gray!70] (0,0) -- plot[smooth] file{rotate_cost_30.data} -- (0.5,0) -- cycle;
        \draw[step=0.1,very thin,color=gray] (0,0) grid (0.5,3);
        \draw[thin,color=black] (0,1) -- (0.5,1);
        \draw[thin,color=black] (0,2) -- (0.5,2);
        \draw[thin,color=black] (0,3) -- (0.5,3);
        \draw (0,1) node[left] {$1$};
        \draw (0,2) node[left] {$2$};
        \draw (0,3) node[left] {$3$};
        \draw (0.5,0) node[below] {$\frac12$};
        \draw[->] (-0.04,0) -- (0.54,0) node[right] {$\RelLength$};
        \draw[->] (0,-0.2) -- (0,3.2) node[above] {$\RelCost[0.3]$};
      \end{tikzpicture}
      \begin{tikzpicture}[xscale=10,yscale=1,domain=0:0.5]
        \fill[color=gray!30] (0,0) -- plot[smooth] file{rotate_cost.data} -- (0.5,0) -- cycle;
        \draw[thick] plot[smooth] file{rotate_cost_50.data};
        \fill[color=gray!70] (0,0) -- plot[smooth] file{rotate_cost_50.data} -- (0.5,0) -- cycle;
        \draw[step=0.1,very thin,color=gray] (0,0) grid (0.5,3);
        \draw[thin,color=black] (0,1) -- (0.5,1);
        \draw[thin,color=black] (0,2) -- (0.5,2);
        \draw[thin,color=black] (0,3) -- (0.5,3);
        \draw (0,1) node[left] {$1$};
        \draw (0,2) node[left] {$2$};
        \draw (0,3) node[left] {$3$};
        \draw (0.5,0) node[below] {$\frac12$};
        \draw[->] (-0.04,0) -- (0.54,0) node[right] {$\RelLength$};
        \draw[->] (0,-0.2) -- (0,3.2) node[above] {$\RelCost[0.5]$};
      \end{tikzpicture}
    \end{center}
    \caption{\label{plot-buffered}
      Expected costs at various buffer sizes.  At a buffer size of
      $10\%$, the expected cost is about $1.73$ moves per element, at $20\%$
      it takes about $1.6$ moves per element, at $30\%$ about $1.49$ moves
      per element, and when the buffer size is $50\%$ and the bottom segment
      is the shorter one (as we may assume by symmetry), the segment will
      fit into the buffer, and we find an expected number of $1.25$ moves
      per element.
    }
  \end{figure}

  \begin{rem}
    Similarly, the function $\RelCostOf[\BufLength]{\TheLength} :=
    \SimpleCostOf{1,\TheLength,\BufLength}$ estimates the number of moves
    per element if the algorithm is using a buffer of size $\BufLength n$.
    As can be seen in Figure~\ref{plot-buffered}, the algorithm takes
    advantage of the buffer even in cases where neither the bottom nor the
    top segment fits into the buffer. The algorithm makes use of the
    buffer as soon as the recursion leads to a subproblem that benefits
    from the use of the buffer.
  \end{rem}

  \section{Asymptotic costs of the block cycle scheme}\label{sec.restglied}
  Lemma~\ref{lemma.moves} relates the move count to the sum remainders
  in the Euclidean algorithm. Consequently, 
  we can analyze the average cost of the algorithm (measured in number
  of moves) using some analytic number theory. 
  Specifically, we refine Theorem~\ref{thm.limit} as follows:
  \begin{thm}\label{thm.main}
    The average cost grows linearly with $\Size$. More precisely,
    for each $\TheEpsilon>0$, we have
    \[
      \AvgCostOf{\Size}
      =
      \TheCostConst
      \Size
      +
      \BigO{\Size^{\frac12+\TheEpsilon}}
    \]
    where
    \[
      \TheCostConst
      \approx 1.85
    \]
    is explicitly known. Specifically, we express $\TheCostConst$ as
    $\TheCostConst=1+4 \TheZetaConst$, where $\TheZetaConst$ is computed in
    two ways as a quickly convergent, double infinite series in
    \eqref{eq.const-c} and \eqref{eq.const-c-alternative} below.
  \end{thm}
  
  \begin{rem}
    In particular, the expected number of moves per element is less
    than $2$. In the average case, the algorithm therefore uses fewer
    moves than the trinity rotation scheme. As we have seen, higher
    moments of the asymptotic distribution also exist. However, as the
    worst case is $3$ moves per element, the higher moments are of little
    practical interest. Numerical approximation yields the standard
    deviation to be about $0.50$ moves per element.
  \end{rem}
  
  The remainder of this section is entirely devoted to the proof of
  Theorem~\ref{thm.main} and the determination of the
  constant~$\TheCostConst$.
  
  As is customary in analytic number theory, we will use the
  $\TheEpsilon>0$-convention that all statements involving $\TheEpsilon$
  are meant to hold for each sufficiently small $\TheEpsilon>0$. A
  typical such statement is $\sum_{d \mid n} 1 = \BigO{n^\TheEpsilon}$.

  The main problem in understanding $\MovesOf{\Size}{\Shift}$ is
  the sum
  \begin{equation}\label{def-mbar}
    \RemainderSumOf{\Size}{\Shift}
    =
    \Remainder[1]+\Remainder[2]+\cdots
  \end{equation}
  of remainders encountered in the Euclidean algorithm.  First,
  consider a run of the Euclidean algorithm for relatively prime numbers
  $\Size=:\Remainder[0]$ and $\Shift=:\Remainder[1]$:
  \begin{align*}
    \Remainder[0] &= \Quotient[0] \Remainder[1] + \Remainder[2] \\
    \Remainder[1] &= \Quotient[1] \Remainder[2] + \Remainder[3] \\
                  &\kern2mm \vdots \\
    \Remainder[\LastIndex-1] &= \Quotient[\LastIndex-1]\Remainder[\LastIndex] + 1 \\
    \Remainder[\LastIndex] &= \Quotient[\LastIndex] \\
    \Remainder[\LastIndex+1] &= 1 \\
    \Remainder[\LastIndex+2] &= 0
                               .
  \end{align*}
  Note $\Quotient[\LastIndex]=\Remainder[\LastIndex]\geq 2$. We define the following shorthand:
  \[
    \ExpansionOf{} = 1\qquad
    \ExpansionOf{\Entry[0]} = \Entry[0]\qquad
    \ExpansionOf{\Entry[0],\Entry[1]} = \Entry[0]\Entry[1] + 1 \qquad
    \ExpansionOf{\Entry[0],\ldots,\Entry[\LastIndex]}
    = \Entry[0]\ExpansionOf{\Entry[1],\ldots,\Entry[\LastIndex]}
    + \ExpansionOf{\Entry[2],\ldots,\Entry[\LastIndex]}
  \]
  and obtain
  \(
    \Remainder[\TheIndex] = \ExpansionOf{\Quotient[\TheIndex],\ldots,\Quotient[\LastIndex]}
  \)
  by induction. In particular, we find
  \(
    \Size = \ExpansionOf{\Quotient[0],\ldots,\Quotient[\LastIndex]}
  \)
  and
  \begin{equation}\label{eq.RemainderSum}
    \RemainderSumOf{\Size}{\Shift}
    =
    \ExpansionOf{\Quotient[1],\ldots,\Quotient[\LastIndex]}
    +
    \ExpansionOf{\Quotient[2],\ldots,\Quotient[\LastIndex]}
    +\cdots+
    \ExpansionOf{\Quotient[\LastIndex-1],\Quotient[\LastIndex]}
    +
    \ExpansionOf{\Quotient[\LastIndex]}
    + 1
    .
  \end{equation}

  Induction also shows
  \[
    \begin{pmatrix}
      \ExpansionOf{\Entry[0],\ldots,\Entry[\LastIndex]}
      &
        \ExpansionOf{\Entry[0],\ldots,\Entry[\LastIndex-1]}
      \\
      \ExpansionOf{\Entry[1],\ldots,\Entry[\LastIndex]}
      &
        \ExpansionOf{\Entry[1],\ldots,\Entry[\LastIndex-1]}
    \end{pmatrix}
    =
    \begin{pmatrix}
      \Entry[0] & 1 \\
      1 & 0
    \end{pmatrix}
    \begin{pmatrix}
      \Entry[1] & 1 \\
      1 & 0
    \end{pmatrix}
    \cdots
    \begin{pmatrix}
      \Entry[\LastIndex] & 1 \\
      1 & 0
    \end{pmatrix}
    .
  \]
  Hence
  \[
    \begin{pmatrix}
      \ExpansionOf{\Entry[0],\ldots,\Entry[\LastIndex]}
      &
        \ExpansionOf{\Entry[0],\ldots,\Entry[\LastIndex-1]}
      \\
      \ExpansionOf{\Entry[1],\ldots,\Entry[\LastIndex]}
      &
        \ExpansionOf{\Entry[1],\ldots,\Entry[\LastIndex-1]}
    \end{pmatrix}
    =
    \begin{pmatrix}
      \ExpansionOf{\Entry[0],\ldots,\Entry[\MidIndex]}
      &
        \ExpansionOf{\Entry[0],\ldots,\Entry[\MidIndex-1]}
      \\
      \ExpansionOf{\Entry[1],\ldots,\Entry[\MidIndex]}
      &
        \ExpansionOf{\Entry[1],\ldots,\Entry[\MidIndex-1]}
    \end{pmatrix}
    \begin{pmatrix}
      \ExpansionOf{\Entry[\MidIndex+1],\ldots,\Entry[\LastIndex]}
      &
        \ExpansionOf{\Entry[\MidIndex+1],\ldots,\Entry[\LastIndex-1]}
      \\
      \ExpansionOf{\Entry[\MidIndex+2],\ldots,\Entry[\LastIndex]}
      &
        \ExpansionOf{\Entry[\MidIndex+2],\ldots,\Entry[\LastIndex-1]}
    \end{pmatrix}
  \]
  and in particular
  \[
    \ExpansionOf{\Entry[0],\ldots,\Entry[\LastIndex]}
    =
    \ExpansionOf{\Entry[0],\ldots,\Entry[\MidIndex]}
    \ExpansionOf{\Entry[\MidIndex+1],\ldots,\Entry[\LastIndex]}
    +
    \ExpansionOf{\Entry[0],\ldots,\Entry[\MidIndex-1]}
    \ExpansionOf{\Entry[\MidIndex+2],\ldots,\Entry[\LastIndex]}
    .
  \]

  Heilbronn~\cite[page~93]{Heilbronn69} has shown that
  \begin{align*}
    \LeftA &= \ExpansionOf{\Entry[0],\ldots,\Entry[\MidIndex]},\\
    \LeftB &= \ExpansionOf{\Entry[\MidIndex+1],\ldots,\Entry[\LastIndex]},\\
    \RightA &= \ExpansionOf{\Entry[0],\ldots,\Entry[\MidIndex-1]},\\
    \RightB &= \ExpansionOf{\Entry[\MidIndex+2],\ldots,\Entry[\LastIndex]}
  \end{align*}
  establishes a one--one correspondence between the sets
  \[
    \SetOf[{
      0\leq\MidIndex<\LastIndex\,,\,
      \Size=\ExpansionOf{\Entry[0],\ldots,\Entry[\LastIndex]},\,
      2 \leq \Entry[0],\,
      2\leq \Entry[\LastIndex]
    }]{
      (\MidIndex,(\Entry[0],\ldots,\Entry[\LastIndex]))
    }
  \]
  and
  \[
    \SetOf[
    \LeftA > \RightA \geq 1,\,
    \LeftB > \RightB \geq 1,\,
    1 = \gcdOf{\LeftA,\RightA} = \gcdOf{\LeftB,\RightB},\,
    \Size = \LeftA\LeftB+\RightA\RightB
    ]{(\LeftA,\LeftB,\RightA,\RightB)}
    .
  \]

  In view of~\eqref{eq.RemainderSum}, we deduce first
  \[
    \Sum[{
      \substack{
        1\leq\Shift\leq\frac{\Size}{2}\\
        1=\gcdOf{\Size,\Shift}
      }
    }]{
      \RemainderSumOf{\Shift}{\Size}
    }
    =
    \Sum[{
      \substack{
        \LeftA > \RightA \geq 1 \\
        \LeftB > \RightB \geq 1 \\
        1 = \gcdOf{\LeftA,\RightA} \\
        1 = \gcdOf{\LeftB,\RightB} \\
        \Size = \LeftA\LeftB+\RightA\RightB
      }
    }]
    \LeftB
    +
    \Sum[{
      \substack{
        1\leq\Shift\leq\frac{\Size}{2}\\
        1=\gcdOf{\Size,\Shift}
      }
    }]{
      \gcdOf{\Size,\Shift}
    }
  \]
  and then
  \begin{equation}\label{eq.heilbron}
    \Sum[{
      \substack{
        1\leq\Shift\leq\frac{\Size}{2}\\
      }
    }]{
      \RemainderSumOf{\Shift}{\Size}
    }
    =
    \Sum[{
      \substack{
        \LeftA > \RightA \geq 1 \\
        \LeftB > \RightB \geq 1 \\
        1 = \gcdOf{\LeftA,\RightA} \\
        \Size = \LeftA\LeftB+\RightA\RightB
      }
    }]
    \LeftB
    +
    \Sum[{
      \substack{
        1\leq\Shift\leq\frac{\Size}{2}\\
      }
    }]{
      \gcdOf{\Size,\Shift}
    }
    .
  \end{equation}

  We call the first term on the right hand side $\InvQuantityOf{\Size}$.
  The second term is $O(n^{1+\TheEpsilon})$, so that
  \begin{equation}\label{eq.invquant}
    \Sum[{
      \substack{
        1\leq\Shift\leq\frac{\Size}{2}\\
      }
    }]{
      \RemainderSumOf{\Shift}{\Size}
    }
    =
    \InvQuantityOf{\Size}
    +
    \BigO{\Size^{1+\TheEpsilon}}
    .
  \end{equation}  
  
  We want to remove the gcd-condition on $\InvQuantityOf{\Size}$,
  and with this in mind, we define
  \[
    \TheQuantityOf{\Size}
    :=
    \Sum[{
      \substack{
        \LeftA > \RightA \geq 1 \\
        \LeftB > \RightB \geq 1 \\
        \Size = \LeftA\LeftB+\RightA\RightB
      }
    }]
    \LeftB
    =
    \Sum[{
      \substack{
        \LeftA > \RightA \geq 1 \\
        \LeftB > \RightB \geq 1 \\
        \Size = \LeftA\LeftB+\RightA\RightB
      }
    }]
    \LeftA
    =\frac12
    \Sum[{
      \substack{
        \LeftA > \RightA \geq 1 \\
        \LeftB > \RightB \geq 1 \\
        \Size = \LeftA\LeftB+\RightA\RightB
      }
    }]
    \Parens{\LeftB+\LeftA}.
  \]
  We can now order the pairs $(\LeftA,\RightA)$ according to their
  gcd (which automatically divides $\Size$) to see that 
  $\TheQuantityOf{\Size} =  \Sum[\TheDivisor\divides\Size]{
    \InvQuantityOf{\TheDivisor}
  }$. 
  We obtain
  \begin{equation}\label{mobius}
    \InvQuantityOf{\Size} =
    \Sum[\TheDivisor\divides\Size]{
      \MoebiusOf{\TheDivisor}
      \TheQuantityOf{\frac{\Size}{\TheDivisor}}
    }
  \end{equation}
  by Möbius inversion.  

  We proceed to analyse $\TheQuantity$. Observe that
  \begin{align*}\label{eq.break-symmetry}
    \TheQuantityOf{\Size}
    &=
      \frac12
      \Sum[\substack{
      \LeftA > \RightA \geq 1 \\
    \LeftB > \RightB \geq 1 \\
    \Size = \LeftA\LeftB+\RightA\RightB
    }]{(\LeftB+\LeftA)}
    =
    \Sum[\substack{
    \LeftA > \RightA \geq 1 \\
    \LeftB > \RightB \geq 1 \\
    \Size = \LeftA\LeftB+\RightA\RightB \\
    \LeftB>\LeftA
    }]{(\LeftB+\LeftA)}
    +
    \BigO{\Size^{1 + \TheEpsilon}}
    \\
    &=
      \Sum[\TheDivisor\divides\Size]{
      \Sum[\substack{
      \LeftA > \RightA \geq 1 \\
    \LeftB > \RightB \geq 1 \\
    \Size = \LeftA\LeftB+\RightA\RightB \\
    \LeftB>\LeftA\\
    \TheDivisor=\gcdOf{\LeftA,\RightA}
    }]{(\LeftB+\LeftA)}
    }
    +
    \BigO{\Size^{1 + \TheEpsilon}}
    =
    \Sum[\TheDivisor\divides\Size]{
    \Sum[\substack{
    \LeftA > \RightA \geq 1 \\
    \LeftB > \RightB \geq 1 \\
    \frac{\Size}{\TheDivisor} = \LeftA\LeftB+\RightA\RightB \\
    \LeftB>\TheDivisor\LeftA\\
    1=\gcdOf{\LeftA,\RightA}
    }]{(\LeftB+\TheDivisor\LeftA)}
    }
    +
    \BigO{\Size^{1 + \TheEpsilon}}
    .
  \end{align*}
  Rewriting the summation conditions yields $\LeftB=
  \frac{\Size}{\TheDivisor\LeftA} -\frac{\RightA\RightB}{\LeftA}$ where
  it is assumed that $\LeftA$ divides $\frac{\Size}{\TheDivisor}
  -\RightA\RightB$. Then, $\LeftB>\RightB$ is equivalent to
  $\frac{\Size}{\TheDivisor(\RightA+\LeftA)}>\RightB$.  Similarly,
  $\LeftB>\TheDivisor\LeftA$ is equivalent to
  $\frac{\Size}{\TheDivisor}-\RightA\RightB > \TheDivisor\LeftA^2$.
  Hence:
  \begin{align*}
    \TheQuantityOf{\Size}
    &=
      \Sum[\TheDivisor\divides\Size]{
        \Sum[\substack{
          \LeftA > \RightA \geq 1 \\
          \frac{\Size}{\TheDivisor(\LeftA+\RightA)}>\RightB \geq 1 \\
          \frac{\Size}{\TheDivisor} \congruent \RightA\RightB \mod \LeftA \\
          \frac{\Size}{\TheDivisor}-\RightA\RightB > \TheDivisor\LeftA^2 \\
          1=\gcdOf{\LeftA,\RightA}
        }]{
          \Parens{\frac{\Size}{\TheDivisor\LeftA} -\frac{\RightA\RightB}{\LeftA}+\TheDivisor\LeftA}
        }
      }
      +
    \BigO{\Size^{1 + \TheEpsilon}}
    .
  \end{align*}

  For the moment, let us fix $\TheDivisor$, $\LeftA$ and $\RightA$ and
  consider the sum
  \[
    \TheQuantityOf{\Size,\TheDivisor,\LeftA,\RightA}
    :=
    \Sum[\substack{
      \frac{\Size}{\TheDivisor(\LeftA+\RightA)}>\RightB \geq 1 \\
      \frac{\Size}{\TheDivisor} \congruent \RightA\RightB \mod \LeftA \\
      \frac{\Size}{\TheDivisor}-\RightA\RightB > \TheDivisor\LeftA^2 \\
    }]{
      \Parens{
        \frac{\Size}{\TheDivisor\LeftA}
        -
        \frac{\RightA\RightB}{\LeftA}+\TheDivisor\LeftA
      }
    }
  \]    
  over $\RightB$. As $\LeftA$ and $\RightA$ are relatively prime, the
  congruence $\frac{\Size}{\TheDivisor} \congruent \RightA\RightB \mod
  \LeftA$ determines the remainder of $\RightB$ modulo $\LeftA$. Let us
  denote by $\TheCongruenceClass$ a representative such that
  $\frac{\Size}{\TheDivisor} \congruent \RightA\RightB \mod \LeftA$ is
  equivalent to $\RightB \congruent \TheCongruenceClass \mod \LeftA$.
  
  Furthermore, we find that
  $\frac{\Size}{\TheDivisor}-\RightA\RightB > \TheDivisor\LeftA^2$ translates into
  $\RightB< \frac{\Size}{\TheDivisor\RightA}-\frac{\TheDivisor\LeftA^2}{\RightA}
  =\frac{\Size-\TheDivisor^2\LeftA^2}{\TheDivisor\RightA}$.
  Thus, $\RightB$ is bounded from above by
  \(
    \UpperBound =
    \UpperBoundOf{\Size,\TheDivisor,\LeftA,\RightA}
    :=
    \MinOf{
      \frac{\Size}{\TheDivisor\Parens{\LeftA+\RightA}},
      \frac{\Size-\TheDivisor^2\LeftA^2}{\TheDivisor\RightA}
    }
  \).

  Consequently
  \begin{align*}
    \TheQuantityOf{\Size,\TheDivisor,\LeftA,\RightA}
    &=
    \Sum[\substack{
      \RightB \congruent \TheCongruenceClass \mod \LeftA \\
      1 \leq \RightB < \UpperBound
    }]{
    \Parens{
    \frac{\Size}{\TheDivisor\LeftA}
    -
    \frac{\RightA\RightB}{\LeftA}+\TheDivisor\LeftA
    }
    }
    =
    \Sum[\substack{
      \RightB \congruent \TheCongruenceClass \mod \LeftA \\
      1 \leq \RightB < \UpperBound
    }]{
      (\ConstantTerm+\LinearTerm\RightB)
    }
  \end{align*}
  where
  \[
    \ConstantTerm
    =  \frac{\Size}{\TheDivisor\LeftA} +\TheDivisor\LeftA
    \qquad\qquad\text{and}\qquad\qquad                   
    \LinearTerm
    = -\frac{\RightA}{\LeftA}
    .
  \]

  We would like to remove the unwieldy summation condition
  $\RightB \congruent \TheCongruenceClass \mod \LeftA$.
  Using the character
  \begin{align*}
    \TheChar \mapcolon \ModRing & \longrightarrow \CCC \\
    \ModRingElement & \mapsto \ExpOf{\frac{2\pi\ImUnit}{\LeftA} \ModRingElement}
  \end{align*}
  we rewrite
  \begin{align*}
    \TheQuantityOf{\Size,\TheDivisor,\LeftA,\RightA}
    &=
      \frac{1}{\LeftA}
      \Sum[\ModRingElement\in\ModRing]{
      \TheCharOf{\ModRingElement\TheCongruenceClass}
      \Sum[1\leq\RightB<\UpperBound]{
      (\ConstantTerm+\LinearTerm\RightB)\TheCharOf{-\ModRingElement\RightB}
      }
      }
    \\
    &=
      \frac{1}{\LeftA}
      \Sum[1\leq\RightB<\UpperBound]{
      (\ConstantTerm+\LinearTerm\RightB)
      }
      +
      \frac{1}{\LeftA}
      \Sum[\ModRingElement\not\congruent 0]{
      \TheCharOf{\ModRingElement\TheCongruenceClass}
      \Sum[1\leq\RightB<\UpperBound]{
      (\ConstantTerm+\LinearTerm\RightB)\TheCharOf{-\ModRingElement\RightB}
      }
      }.
  \end{align*}
  The main term will come from the first term on the right hand
  side. Note that
  \[
    \AbsValueOf{
      \frac{1}{\LeftA}
      (\UpperBound\ConstantTerm+\frac{\UpperBound^2}{2}\LinearTerm)
      \,\,-\,\,
      \frac{1}{\LeftA}
      \Sum[1\leq\RightB<\UpperBound]{
        (\ConstantTerm+\LinearTerm\RightB)
      }
    }
    \leq
    \AbsValueOf{\ConstantTerm}
    +
    \AbsValueOf{\LinearTerm\UpperBound}.
  \]
  Thus we have
    $ \TheQuantityOf{\Size,\TheDivisor,\LeftA,\RightA} =
    \TheQuantityOf[1]{\Size,\TheDivisor,\LeftA,\RightA} +
    \TheQuantityOf[2]{\Size,\TheDivisor,\LeftA,\RightA} +
    \TheQuantityOf[3]{\Size,\TheDivisor,\LeftA,\RightA}$
  where
  \begin{align*}
    \TheQuantityOf[1]{\Size,\TheDivisor,\LeftA,\RightA}
    &=
      \frac{1}{\LeftA}
      (\UpperBound\ConstantTerm+\frac{\UpperBound^2}{2}\LinearTerm), \quad\quad 
      \AbsValueOf{\TheQuantityOf[2]{\Size,\TheDivisor,\LeftA,\RightA}}
      \leq \AbsValueOf{\ConstantTerm}+ \AbsValueOf{\LinearTerm\UpperBound},
    \\
      \TheQuantityOf[3]{\Size,\TheDivisor,\LeftA,\RightA}
    &=
      \frac{1}{\LeftA}
      \Sum[\ModRingElement\not\congruent 0]{
      \TheCharOf{\ModRingElement\TheCongruenceClass}
      \Sum[1\leq\RightB<\UpperBound]{
      (\ConstantTerm+\LinearTerm\RightB)\TheCharOf{-\ModRingElement\RightB}
      }
      }
      .
  \end{align*}
  Thus
  \[
    \TheQuantityOf{\Size}
    =
    \TheQuantityOf[1]{\Size}
    +
    \TheQuantityOf[2]{\Size}
    +
    \TheQuantityOf[3]{\Size}
  \]
  where
  \[
    \TheQuantityOf[\AltIndex]{\Size}
    =
    \Sum[\TheDivisor\divides\Size]{
      \Sum[\substack{
        \LeftA>\RightA\geq 1 \\
        1 = \gcdOf{\LeftA,\RightA} \\
        1 < \UpperBoundOf{\Size,\TheDivisor,\LeftA,\RightA}
    }]{
      \TheQuantityOf[\AltIndex]{\Size,\TheDivisor,\LeftA,\RightA}
    }
  }
  .
  \]

  \begin{lemma}\label{lemma.g-two}
    We have
    \(
      \TheQuantityOf[2]{\Size}
      =
      \BigO{\Size^{\frac32 + \TheEpsilon}}
      .
    \)
  \end{lemma}
  \begin{proof}
    Note that $\UpperBound\geq1$ implies $\Size > \TheDivisor^2\LeftA^2$. I.e.,
    $\LeftA < \frac{\SqrtOf{\Size}}{\TheDivisor}$ is a necessary condition for
    $\UpperBound \geq 1$. Moreover,  we have
    $\UpperBound\leq\frac{\Size}{\TheDivisor\LeftA}$. 
    Thus
    \begin{displaymath}
    \begin{split}
      \AbsValueOf{
      \TheQuantityOf[2]{\Size}}
      &
        \leq
        \Sum[\TheDivisor\divides\Size]{
        \Sum[\substack{
          1 \leq \RightA < \LeftA < \frac{\SqrtOf{\Size}}{\TheDivisor} \\
        }]{
          \AbsValueOf{\ConstantTerm} +   \AbsValueOf{\LinearTerm\UpperBound}
      }}
      \leq
      \Sum[\TheDivisor\divides\Size]{
        \Sum[\substack{
          1 \leq \RightA < \LeftA < \frac{\SqrtOf{\Size}}{\TheDivisor}
        }]{
          \frac{\Size}{\TheDivisor\LeftA}+\TheDivisor\LeftA
        +
        \frac{\RightA \Size}{\TheDivisor\LeftA^2 }
        }
      }\\
      &
        \leq
        \Sum[\TheDivisor\divides\Size]{
        \Sum[\substack{
        1 \leq \LeftA < \frac{\SqrtOf{\Size}}{\TheDivisor}
        }]{
        \frac{\Size}{\TheDivisor}+\TheDivisor\LeftA^2
        +
        \frac{\Size}{\TheDivisor}
        }
        }
        \leq
        \Sum[\TheDivisor\divides\Size]{
        \frac{\SqrtOf{\Size}}{\TheDivisor}
        \left(
        \frac{\Size}{\TheDivisor}
        +
        \frac{\Size}{\TheDivisor}
        +
        \frac{\Size}{\TheDivisor}
        \right)
        }
          =
        \BigO{\Size^{\frac32 + \TheEpsilon}}.
    \end{split}
  \end{displaymath}
  \end{proof}

  \begin{obs}\label{obs.exp-sum}
    There are two obvious metrics on the unit circle in the complex plane:
    the intrinsic metric using arc length and the euclidean metric induced
    by the embedding. Both metrics are bi-Lipschitz equivalent.
    More precisely, for $\TheAngle\in[-\pi,\pi]$, we have
    \[
      \frac{2}{\pi}\AbsValueOf{\TheAngle}
      \leq
      \AbsValueOf{\ExpOf{\ImUnit\TheAngle}-1}
      \leq \AbsValueOf{\TheAngle}, 
    \]
    so
    \[
      \AbsValueOf{
        \sum_{\TheIndex=\BotBound}^{\TopBound-1}
        \ExpOf{\ImUnit\TheAngle\TheIndex}
      }
      =
      \AbsValueOf{ \frac{
          \ExpOf{\ImUnit\TheAngle\BotBound}-\ExpOf{\ImUnit\TheAngle\TopBound}
        }{
          1 - \ExpOf{\ImUnit\TheAngle}
        }}
      \leq\frac{2}{
        \AbsValueOf{\ExpOf{\ImUnit\TheAngle}-1}
      } 
      \leq
      \AbsValueOf{\frac{\pi}{\TheAngle}}
    \]
    for $\TheAngle \not= 0$. Similarly, again for $\TheAngle \not= 0$, from
    \begin{align*}
      \sum_{\TheIndex=1}^{\TopBound-1}
      \TheIndex\ExpOf{\ImUnit\TheAngle\TheIndex}
      &=
        \sum_{\BotBound=1}^{\TopBound-1}
        \sum_{\TheIndex=\BotBound}^{\TopBound-1}
        \ExpOf{\ImUnit\TheAngle\TheIndex}
        =
        \frac{
        \ExpOf{\ImUnit\TheAngle}-\ExpOf{\ImUnit\TheAngle\TopBound}
        }{
        (1 - \ExpOf{\ImUnit\TheAngle})^2
        }
        -
        \frac{  
        \Parens{\TopBound-1}\ExpOf{\ImUnit\TheAngle\TopBound}
        }{
        1 - \ExpOf{\ImUnit\TheAngle}
        }
    \end{align*}
    we deduce
    \[
      \AbsValueOf{
        \sum_{\TheIndex=1}^{\TopBound-1}
        \TheIndex\ExpOf{\ImUnit\TheAngle\TheIndex}
      }
      \leq
      \frac{\pi^2}{2\AbsValueOf{\TheAngle}^2}
      +
      \frac{\Parens{\TopBound-1}\pi}{2\AbsValueOf{\TheAngle}}
      .
    \]
  \end{obs}
  
  \begin{lemma}\label{lemma.g-three}
    We have
    \(
      \TheQuantityOf[3]{\Size}
      =
      \BigO{\Size^{\frac32 + \TheEpsilon}}
      .
    \)
  \end{lemma}
  \begin{proof}
    Recall that
    \[
      \TheQuantityOf[3]{\Size,\TheDivisor,\LeftA,\RightA}
      =
      \frac{1}{\LeftA}
      \Sum[\ModRingElement\not\congruent 0]{
        \TheCharOf{\ModRingElement\TheCongruenceClass}
        \Sum[1\leq\RightB<\UpperBound]{
          (\ConstantTerm+\LinearTerm\RightB)\TheCharOf{-\ModRingElement\RightB}. 
        }
      }      
    \]
    Choosing the representative for $0 \not\equiv\ModRingElement\in\ModRing$
    between $-\frac{\LeftA}{2}$ and $\frac{\LeftA}{2}$,
    we deduce from Observation~\ref{obs.exp-sum} that
    \begin{align*}
      \AbsValueOf{
      \frac{1}{\LeftA}
      \Sum[1\leq\RightB<\UpperBound]{
      \Parens{\ConstantTerm+\LinearTerm\RightB}
      \ExpOf{-2\pi\ImUnit\frac{\ModRingElement}{\LeftA}\RightB}
      }
      }
      &\leq
        \frac{1}{|\LeftA|}\Big(
        \ConstantTerm\pi\frac{|\LeftA|}{2\pi\AbsRingElement}
        +
        \AbsValueOf{\LinearTerm}
        \frac{\pi^2}{2}\Parens{
        \frac{\LeftA}{2\pi\AbsRingElement}
        }^2
        +       \AbsValueOf{\LinearTerm}
        \frac{\UpperBound\pi}{2}\frac{|\LeftA|}{2\pi\AbsRingElement} \Big)\\
      &\leq
      \Parens{
        \frac{\Size}{\TheDivisor|\LeftA|} +\TheDivisor|\LeftA|
      }\frac{1}{2\AbsRingElement}
      +
      \frac{\RightA}{8\AbsRingElement^2}
      +
        \frac{\Size}{\TheDivisor\Parens{|\LeftA|+\RightA}}\frac{\RightA}{4\AbsRingElement|\LeftA|}
        .
    \end{align*}
    Hence
    \begin{align*}
      \AbsValueOf{\TheQuantityOf[3]{\Size,\TheDivisor,\LeftA,\RightA}}
      &\leq
      2
      \sum_{\ModRingElement=1}^{\LeftA/2}
      \Parens{
        \frac{\Size}{\TheDivisor\LeftA} +\TheDivisor\LeftA
      }\frac{1}{2\AbsRingElement}
      +
      2
      \sum_{\ModRingElement=1}^{\LeftA/2}
      \frac{\RightA}{8\AbsRingElement^2}
      +
      2
      \sum_{\ModRingElement=1}^{\LeftA/2}
      \frac{\Size}{\TheDivisor\Parens{\LeftA+\RightA}}\frac{\RightA}{4\AbsRingElement\LeftA}
      \\
      &\leq
      \Parens{
        \frac{\Size}{\TheDivisor\LeftA} +\TheDivisor\LeftA
      }\LogOf{\LeftA}
      +
      \frac{\pi^2}{24}\RightA
      +
        \frac{\Size}{\TheDivisor\Parens{\LeftA+\RightA}}
        \frac{\RightA}{2\LeftA}\LogOf{\LeftA}
        .
    \end{align*}
    From
    $\UpperBoundOf{\Size,\TheDivisor,\LeftA,\RightA}
    \leq
    \frac{\SqrtOf{\LeftA}}{\TheDivisor}$
    we deduce
    \[
      \AbsValueOf{\TheQuantityOf[3]{\Size}}
      \leq
      \Sum[\TheDivisor\divides\Size]{
        \Sum[{
          1\leq\RightA<\LeftA\leq\frac{\SqrtOf{\Size}}{\TheDivisor}
        }]{
          \AbsValueOf{\TheQuantityOf[3]{\Size,\TheDivisor,\LeftA,\RightA}}
        }
      }
      .
    \]
    Splitting the sum according to the bound for
    $\AbsValueOf{\TheQuantityOf[3]{\Size,\TheDivisor,\LeftA,\RightA}}$, we find
    the following estimates:
    \[
      \Sum[\TheDivisor\divides\Size]{
        \Sum[{
          1\leq\RightA<\LeftA\leq\frac{\SqrtOf{\Size}}{\TheDivisor}
        }]{
          \Parens{
            \frac{\Size}{\TheDivisor\LeftA} +\TheDivisor\LeftA
          }\LogOf{\LeftA}          
        }
      }
      \leq
      \Sum[\TheDivisor\divides\Size]{
        \Sum[{
          1\leq\LeftA\leq\frac{\SqrtOf{\Size}}{\TheDivisor}
        }]{
          \Parens{
            \frac{\Size}{\TheDivisor} +\frac{\Size}{\TheDivisor}
          }\LogOf{\LeftA}          
        }
      }
      =\BigO{\Size^{\frac32+\TheEpsilon}}
      ,
    \]
    \[
      \Sum[\TheDivisor\divides\Size]{
        \Sum[{
          1\leq\RightA<\LeftA\leq\frac{\SqrtOf{\Size}}{\TheDivisor}
        }]{
          \frac{\pi^2}{24}\RightA
        }
      }
      \leq
      \Sum[\TheDivisor\divides\Size]{
        \frac{\Size}{\TheDivisor^2}
        \frac{\pi^2}{24}
        \frac{\SqrtOf{\Size}}{\TheDivisor}
      }
      =\BigO{\Size^{\frac32+\TheEpsilon}}
      ,
    \]
    \[
      \Sum[\TheDivisor\divides\Size]{
        \Sum[{
          1\leq\RightA<\LeftA\leq\frac{\SqrtOf{\Size}}{\TheDivisor}
        }]{
          \frac{\Size}{\TheDivisor\Parens{\LeftA+\RightA}}\frac{\RightA}{2\LeftA}\LogOf{\LeftA}
        }
      }
      \leq
      \Sum[\TheDivisor\divides\Size]{
        \Sum[{
          1\leq\LeftA\leq\frac{\SqrtOf{\Size}}{\TheDivisor}
        }]{
          \frac{\Size\LeftA}{2\TheDivisor\Parens{\LeftA+1}}\LogOf{\LeftA}
        }
      }
      =\BigO{\Size^{\frac32+\TheEpsilon}}
      .
    \]
    The claim follows.
  \end{proof}

  \begin{lemma}\label{lemma.g-one}
    We have
    \[
      \TheQuantityOf[1]{\Size}
      =
      \TheZetaConst
      \Size^2
      \Sum[\TheDivisor\divides\Size]{
        \frac{1}{\TheDivisor^2}
      }
      +
      \BigO{\Size^{\frac32 + \TheEpsilon}}
    \]
    where
    \begin{equation}\label{eq.const-c}
      \TheZetaConst
      =
      \Sum[\substack{
        \LeftA>\RightA\geq 1 \\
        1 = \gcdOf{\LeftA,\RightA} \\
      }]{
        \frac{
          2\LeftA+\RightA
        }{
          2\LeftA^2\Parens{\LeftA+\RightA}^2
        }
      }
      .
    \end{equation}
  \end{lemma}
  \begin{proof}
    Recall
    \[
      \TheQuantityOf[1]{\Size}
      =
      \Sum[\TheDivisor\divides\Size]{
        \Sum[\substack{
          \LeftA>\RightA\geq 1 \\
          1 = \gcdOf{\LeftA,\RightA} \\
          1 < \UpperBoundOf{\Size,\TheDivisor,\LeftA,\RightA}
        }]{
          \TheQuantityOf[1]{\Size,\TheDivisor,\LeftA,\RightA}
        }
      }
      =
      \Sum[\TheDivisor\divides\Size]{
        \Sum[\substack{
          \LeftA>\RightA\geq 1 \\
          1 = \gcdOf{\LeftA,\RightA} \\
          1 < \UpperBoundOf{\Size,\TheDivisor,\LeftA,\RightA}
        }]{
          \frac{1}{\LeftA}\Parens{\ConstantTerm\UpperBound+\LinearTerm\frac{\UpperBound^2}{2}}
        }
      }
    \]
    and note that
    \[
      \UpperBoundOf{\Size,\TheDivisor,\LeftA,\RightA}
      =
      \MinOf{
        \frac{\Size}{\TheDivisor\Parens{\LeftA+\RightA}}
        ,
        \frac{\Size-\TheDivisor^2\LeftA^2}{\TheDivisor\RightA}
      }
      =
      \begin{cases}
        \frac{\Size}{\TheDivisor\Parens{\LeftA+\RightA}}
        &
          \text{if\ }
          \RightA \leq \frac{\Size}{\TheDivisor^2\LeftA} - \LeftA
        \\
        \frac{\Size-\TheDivisor^2\LeftA^2}{\TheDivisor\RightA}
        &
            \text{if\ }
          \RightA \geq \frac{\Size}{\TheDivisor^2\LeftA} - \LeftA
          .
      \end{cases}
    \]
    We split the sum
    \[
      \TheQuantityOf[1]{\Size}
      =
      \Sum[\TheDivisor\divides\Size]{
        \BulkQuantityOf{\Size,\TheDivisor}
      }
      +
      \Sum[\TheDivisor\divides\Size]{
        \SmallQuantityOf{\Size,\TheDivisor}
      }
    \]
    with
    \begin{align*}
      \BulkQuantityOf{\Size,\TheDivisor}
      &:=
        \Sum[\substack{
        \LeftA>\RightA\geq 1 \\
      1 = \gcdOf{\LeftA,\RightA} \\
      \RightA < \frac{\Size}{\TheDivisor^2\LeftA} - \LeftA \\
      1 < \frac{\Size}{\TheDivisor\Parens{\LeftA+\RightA}}
        }]{
          \frac{1}{\LeftA}\Parens{\ConstantTerm\UpperBound+\LinearTerm\frac{\UpperBound^2}{2}}
      }
      , \quad 
      \SmallQuantityOf{\Size,\TheDivisor}
        :=
        \Sum[\substack{
        \LeftA>\RightA\geq 1 \\
      1 = \gcdOf{\LeftA,\RightA} \\
      \RightA \geq \frac{\Size}{\TheDivisor^2\LeftA} - \LeftA \\
      1 < \frac{\Size-\TheDivisor^2\LeftA^2}{\TheDivisor\RightA}
      }]{
      \frac{1}{\LeftA}\Parens{\ConstantTerm\UpperBound+\LinearTerm\frac{\UpperBound^2}{2}}
      }
      .
    \end{align*}
    Unraveling the definitions of $\ConstantTerm$ and $\LinearTerm$,
    a straightforward computation yields
    \begin{align*}
      \SmallQuantityOf{\Size,\TheDivisor}
      & =
      \Sum[\substack{
      \LeftA>\RightA\geq 1 \\
      1 = \gcdOf{\LeftA,\RightA} \\
      \RightA \geq \frac{\Size}{\TheDivisor^2\LeftA} - \LeftA \\
      1 < \frac{\Size-\TheDivisor^2\LeftA^2}{\TheDivisor\RightA}
      }]{
      \Parens{
      \frac{\Size^2}{2\TheDivisor^2\LeftA^2\RightA}
      +
      \frac{\Size}{\RightA}
      -
      \frac{3\TheDivisor^2\LeftA^2}{2\RightA}
      }}.
    \end{align*}
    Since $\LeftA$ and $\RightA$ are bounded from above by
    $\SqrtOf{\Size}/\TheDivisor$ and $2\LeftA^2$ is bounded from below
    by $\frac{\Size}{\TheDivisor^2}$, we find that
    \(
      \SmallQuantityOf{\Size,\TheDivisor}
      =
      \BigO{\Size^{\frac32}}
    \)
    whence
    \[
      \Sum[\TheDivisor\divides\Size]{
        \SmallQuantityOf{\Size,\TheDivisor}
      }
      =
      \BigO{\Size^{\frac32 + \TheEpsilon}}.
    \]

    In order to analyze
    $\BulkQuantityOf{\Size,\TheDivisor}$, we first simplify the summation
    conditions.  From $\RightA < \frac{\Size}{\TheDivisor^2\LeftA} -
    \LeftA$, we deduce first $\LeftA+\RightA <
    \frac{\Size}{\TheDivisor^2\LeftA}$ and then
    $\TheDivisor\LeftA<\frac{\Size}{\TheDivisor\Parens{\LeftA+\RightA}}$. In
    particular, the condition $1 <
    \frac{\Size}{\TheDivisor\Parens{\LeftA+\RightA}}$ is redundant in view
    of $\LeftA>\RightA\geq 1$ and note that $\LeftA^2 \leq
    \Size/\TheDivisor^2$. Thus
    \begin{align*}
      \BulkQuantityOf{\Size,\TheDivisor}
      & =
        \Sum[\substack{
        \LeftA>\RightA\geq 1 \\
      1 = \gcdOf{\LeftA,\RightA} \\
      \RightA < \frac{\Size}{\TheDivisor^2\LeftA} - \LeftA
        }]{
      \frac{1}{\LeftA}
      \Parens[\Bigg]{
        \ConstantTerm\frac{\Size}{\TheDivisor\Parens{\LeftA+\RightA}}
      +\frac{\LinearTerm}{2}\Parens{\frac{\Size}{\TheDivisor\Parens{\LeftA+\RightA}}}^2
      }
      }
      \\
      & =
        \Sum[\substack{
        \LeftA>\RightA\geq 1 \\
      1 = \gcdOf{\LeftA,\RightA} \\
      \RightA < \frac{\Size}{\TheDivisor^2\LeftA} - \LeftA
        }]{
      \Parens{
      \frac{\Size^2}{\TheDivisor^2\LeftA^2\Parens{\LeftA+\RightA}}
      +
      \frac{\Size}{\Parens{\LeftA+\RightA}}
      -\frac{\Size^2\RightA}{2\LeftA^2\TheDivisor^2\Parens{\LeftA+\RightA}^2}
      }
      }
      \\
      & =
        \Sum[\substack{
        \LeftA>\RightA\geq 1 \\
      1 = \gcdOf{\LeftA,\RightA} \\
      \RightA < \frac{\Size}{\TheDivisor^2\LeftA} - \LeftA
        }]{
      \Parens{
      \frac{\Size^2}{\TheDivisor^2\LeftA^2\Parens{\LeftA+\RightA}}
      -\frac{\Size^2\RightA}{2\LeftA^2\TheDivisor^2\Parens{\LeftA+\RightA}^2}
      }
      }
      +
      \BigO{\Size^{\frac32}}
      \\
      & =
        \frac{\Size^2}{\TheDivisor^2}
        \Sum[\substack{
        \LeftA>\RightA\geq 1 \\
      1 = \gcdOf{\LeftA,\RightA} \\
      \RightA < \frac{\Size}{\TheDivisor^2\LeftA} - \LeftA
        }]{
      \Parens{
      \frac{2\LeftA+\RightA}{2\LeftA^2\Parens{\LeftA+\RightA}^2}
      }
      }
      +
      \BigO{\Size^{\frac32}}
    \end{align*}
    and
    \begin{equation}\label{eq.bulk}
      \Sum[\TheDivisor\divides\Size]{
        \BulkQuantityOf{\Size,\TheDivisor}
      }
      =
      \Sum[\TheDivisor\divides\Size]{
        \frac{\Size^2}{\TheDivisor^2}
        \Sum[\substack{
          \LeftA>\RightA\geq 1 \\
          1 = \gcdOf{\LeftA,\RightA} \\
          \RightA < \frac{\Size}{\TheDivisor^2\LeftA} - \LeftA
        }]{
          \Parens{
            \frac{2\LeftA+\RightA}{2\LeftA^2\Parens{\LeftA+\RightA}^2}
          }
        }
      }
      +
      \BigO{\Size^{\frac32 + \TheEpsilon}}
      .
    \end{equation}
    We want to remove the truncation
    $\RightA < \frac{\Size}{\TheDivisor^2\LeftA} - \LeftA$
    and therefore estimate the tail.
    Note that
    $\RightA\geq\frac{\Size}{\TheDivisor^2\LeftA} - \LeftA$
    together with $\LeftA>\RightA$ implies
    $2\TheDivisor^2\LeftA^2>\Size$, i.e.,
    $\LeftA>\frac{\SqrtOf{\Size}}{\SqrtOf{2}\TheDivisor}$.
    Thus we have
    \begin{align*}
      \Sum[\TheDivisor\divides\Size]{
      \frac{\Size^2}{\TheDivisor^2}
      \Sum[\substack{
      \LeftA>\RightA\geq 1 \\
      1 = \gcdOf{\LeftA,\RightA} \\
      \RightA \geq \frac{\Size}{\TheDivisor^2\LeftA} - \LeftA
      }]{
      \frac{2\LeftA+\RightA}{2\LeftA^2\Parens{\LeftA+\RightA}^2}   
      }
      }
      &
        \leq
        \Sum[\TheDivisor\divides\Size]{
        \frac{\Size^2}{\TheDivisor^2}
        \Sum[\substack{
        \LeftA>\frac{\SqrtOf{\Size}}{\SqrtOf{2}\TheDivisor}\\
      1 \leq \RightA <\LeftA
      }]{
      \frac{1}{\LeftA^3}
      }
      }
        \leq
        \Sum[\TheDivisor\divides\Size]{
        \Size^2
        \Sum[\substack{
        \LeftA>\frac{\SqrtOf{\Size}}{\SqrtOf{2}\TheDivisor}\\
      }]{
      \frac{1}{\TheDivisor^2\LeftA^2}
      }
      }
      \\
       &
         \leq
         \Sum[\TheDivisor\divides\Size]{
         \Size^2
         \Sum[\substack{
         \AltIndex>\frac{\SqrtOf{\Size}}{\SqrtOf{2}}\\
      }]{
      \frac{1}{\AltIndex^2}
      }
      }
        =
        \BigO{\Size^{\frac32 + \TheEpsilon}}
        .
    \end{align*}
    Combining the previous estimates, it follows that 
    \begin{align*}
      \TheQuantityOf[1]{\Size}
      &
        =
      \Sum[\TheDivisor\divides\Size]{
        \frac{\Size^2}{\TheDivisor^2}
        \Sum[\substack{
          \LeftA>\RightA\geq 1 \\
          1 = \gcdOf{\LeftA,\RightA}
        }]{
          \frac{2\LeftA+\RightA}{2\LeftA^2\Parens{\LeftA+\RightA}^2}
        }
      }
      +
      \BigO{\Size^{\frac32 + \TheEpsilon}}
    \end{align*}
    as claimed.
  \end{proof}

  \begin{proof}[Proof of Theorem~\ref{thm.main}]
    Recall that
    \(
      \TheQuantityOf{\Size}
      =
      \TheQuantityOf[1]{\Size}
      +
      \TheQuantityOf[2]{\Size}
      +
      \TheQuantityOf[3]{\Size}
    \).
    In view of Lemma~\ref{lemma.g-two}, Lemma~\ref{lemma.g-three}, and
    Lemma~\ref{lemma.g-one}, we infer
    \[
      \TheQuantityOf{\Size}
      =
      \TheZetaConst\Size^2
      \Sum[\TheDivisor\divides\Size]{
        \frac{1}{\TheDivisor^2}
      }
      +
      \BigO{\Size^{\frac32 + \TheEpsilon}}, \quad  \TheZetaConst
      =
      \Sum[\substack{
        \LeftA>\RightA\geq 1 \\
        1 = \gcdOf{\LeftA,\RightA}
      }]{
        \frac{2\LeftA+\RightA}{2\LeftA^2\Parens{\LeftA+\RightA}^2}
      }.
    \]
    It remains to recall \eqref{eq.invquant} and \eqref{mobius} and observe that 
    \[
      \Sum[\TheDivisor\divides\Size]{ \TheDivisor^2}
      =
      \Sum[\AltDivisor\divides\Size]{\frac{\Size^2}{\AltDivisor^2}}
      =
      \Size^2\Sum[\AltDivisor\divides\Size]{\frac{1}{\AltDivisor^2}}
    \]
    implies by Möbius inversion
    \[
      \Size^2
      =
      \Sum[\TheDivisor\divides\Size]{
        \MoebiusOf{\TheDivisor}
        \frac{\Size^2}{\TheDivisor^2}
        \Sum[\AltDivisor\divides\frac{\Size}{\TheDivisor}]{
          \frac{1}{\AltDivisor^2}
        }
      }
      ,
    \]
    and so 
    \[
      \Sum[{
        \substack{
          1\leq\Shift\leq\frac{\Size}{2}\\
        }
      }]{
        \RemainderSumOf{\Shift}{\Size}
      }
      =  \Sum[\TheDivisor\divides\Size]{
        \MoebiusOf{\TheDivisor}
        \TheQuantityOf{\frac{\Size}{\TheDivisor}}
        +
        \BigO{\Size^{ 1  +\TheEpsilon}}} =  \TheConst \Size^2
      +
      \BigO{\Size^{\frac32 + \TheEpsilon}}.
    \]
    Using \eqref{def-mbar},
    Lemma~\ref{lemma.moves} and \eqref{relation}, we conclude
    \[
      \Sum[\Shift\leq\frac{\Size}{2}]{
        \MovesOf{\Size}{\Shift}
      }
      =
      \frac{\Size^2}{2}
      +
      2
      \Sum[\Shift\leq\frac{\Size}{2}]{
        \RemainderSumOf{\Size}{\Shift}
      }
      +
      \BigO{\Size^{1+\TheEpsilon}}
      =
      \Parens{\frac12 + 2\TheZetaConst}\Size^2
      +
      \BigO{\Size^{\frac32+\TheEpsilon}}
      .
    \]
    Since $\MovesOf{\Size}{\Shift}=\MovesOf{\Size}{\Size-\Shift}$,
    we finally obtain
    \[ 
      \AvgCostOf{\Size}
      =
      \frac{2}{\Size}
      \Sum[\Shift\leq\frac{\Size}{2}]{
        \MovesOf{\Size}{\Shift}
      }
      =
      \Parens{1+4\TheZetaConst}\Size + \BigO{\Size^{\frac12 + \TheEpsilon}}
      ,
    \]
    which proves the claim with the constant
    \(
      \TheCostConst = 1 + 4 \TheZetaConst
    \).
  \end{proof}

  \begin{rem}
    A run of the Euclidean algorithm for the pair $(\Size,\Shift)$ with
    $\Size>\Shift\geq\frac{\Size}{2}$ will lead to the first remainder $\Shift$ and
    then reproduce the same remainders as a run of $(\Size,\Size-\Shift)$. Hence, in this
    case:
    \[
      \RemainderSumOf{\Size}{\Shift} =
      \Shift + \RemainderSumOf{\Size}{\Size-\Shift}
    \]
    From this observation, it follows that the remainder sum in the
    Euclidean algorithm averaged over the range $1\leq\Shift\leq\Size$
    is
    \[
      \Parens{\frac38 + 2\TheZetaConst} \Size + \BigO{\Size^{\frac12 + \TheEpsilon}}. 
    \]
  \end{rem}

  \begin{rem}\label{rem.rewrite-c}
    We can rewrite the constant $\TheZetaConst$ in a more palatable way.
    We have
    \begin{align*}
      \ZetaOf{3}
      \TheZetaConst
      &=
        \Parens{
        \Sum[\TheDivisor]{
        \frac{1}{\TheDivisor^3}
        }
        }
        \Sum[\substack{
        \LeftA>\RightA\geq 1 \\
      1 = \gcdOf{\LeftA,\RightA}
      }]{
      \frac{1}{2\RightA}
      \Parens{
      \frac{1}{\LeftA^2}-\frac{1}{\Parens{\LeftA+\RightA}^2}
      }
      }
      \\
      &=
      \Sum[\substack{
        \LeftA>\RightA\geq 1
        }]{
      \frac{1}{2\RightA}
      \Parens{
      \frac{1}{\LeftA^2}-\frac{1}{\Parens{\LeftA+\RightA}^2}
      }
        }
        =
        \frac{1}{2}
        \ZetaOf{3}
        -
        \frac{1}{2}
        \Sum[\substack{
        \LeftA>\RightA\geq 1
        }]{
        \frac{1}{\RightA\Parens{\LeftA+\RightA}^2}
        }
        ,
    \end{align*}
    where in the penultimate step we used Euler's formula for multiple
    zeta values, see e.g.\
    \cite[{p.\ 509, last display with $k=3$}]{MR1341859}. Thus, we find
    \begin{equation}\label{eq.const-c-alternative}
      \TheZetaConst
      =
      \frac{1}{2}
      -
      \frac{1}{2\ZetaOf{3}}
      \Sum[\substack{
        \LeftA>\RightA\geq 1
      }]{
        \frac{1}{\RightA\Parens{\LeftA+\RightA}^2}
      }
      .
    \end{equation}
    We note that any truncation of the infinite sum yields an upper bound
    for $\TheZetaConst$ and hence for $\TheCostConst$.
  \end{rem}

  \section{Implementations and benchmarks}\label{sec.benchmarks}
  In order to compare the run times of the block cycle scheme to the
  block swap algorithms, to triple reversal, and to trinity rotation, we
  present \verb-C++- implementation of each as generic algorithms
  operating on random access iterators.

  \subsection*{Implementation of the block cycle algorithm}
  In our implementation, we use a large buffer (governed by the constant
  \verb+BUFFER_SIZE+) to end the recursion and a small buffer (governed
  by the constant \verb+BATCH_SIZE+) for batch processing of cycle
  permutations. In the benchmarks, the large buffer accommodates
  $256$ array items whereas the small buffer is $32$ bytes.
  
  Also, we implement an early exit when the left segment
  and the right segment are of equal size, in which case, we just call
  \verb+std::swap_ranges()+.

\begin{lstlisting}
constexpr static struct block_cycle_t {

  template < typename Iterator >
  Iterator operator () ( Iterator from, Iterator middle, Iterator to ) const {
    std::size_t l_size = middle - from;
    std::size_t r_size = to - middle;
      
    using value_type = std::decay_t< decltype( *from ) >;
    constexpr std::size_t const buffer_size = BUFFER_SIZE;
    constexpr std::size_t const batch_size =
      std::max( std::size_t( 1 ), BATCH_SIZE / sizeof( value_type ) );
    static_assert( batch_size <= buffer_size );
    thread_local std::array< value_type, buffer_size > buffer;

    while ( l_size > 0 && r_size > 0 ) {
 
      if ( l_size < r_size  ) {
        if ( l_size <= buffer_size ) {
          std::move( from, middle, buffer.begin() );
          std::move( middle, to, from );
          std::move( buffer.begin(), buffer.begin() + l_size, to - l_size );
          break;
        }

        std::size_t i = 0;
        for ( ; i < l_size - batch_size; i += batch_size ) {
          std::move( from + i, from + i + batch_size, buffer.begin() );
          std::size_t shift = 0;
          for ( ; shift <= r_size - l_size; shift += l_size ) {
            std::move( from + i + shift + l_size,
                       from + i + shift + l_size + batch_size,
                       from + i + shift );
          }
          std::move( buffer.begin(),
                     buffer.begin() + batch_size,
                     from + i + shift );
        }
        std::size_t n_remaining = l_size - i;
        std::size_t shift = 0;
        std::move( from + i, from + i + n_remaining, buffer.begin() );
        for ( ; shift + l_size <= r_size; shift += l_size ) {
          std::move( from + i + shift + l_size,
                     from + i + shift + l_size + n_remaining,
                     from + i + shift );
        }          
        std::move( buffer.begin(),
                   buffer.begin() + n_remaining,
                   from + i + shift );
        from   += shift;
        middle += shift;
        r_size -= shift;
        continue;
      }

      if ( r_size < l_size ) {
        if ( r_size <= buffer_size ) {
          std::move( middle, to, buffer.begin() );
          std::move_backward( from, middle, to );
          std::move( buffer.begin(), buffer.begin() + r_size, from );
          break;
        }

        std::size_t i = 0;
        for ( ; i < r_size - batch_size; i += batch_size ) {
          std::move( middle + i, middle + i + batch_size, buffer.begin() );
          std::size_t shift = 0;
          for ( ; shift <= l_size - r_size; shift += r_size ) {
            std::move( middle + i - shift - r_size,
                       middle + i -shift - r_size + batch_size,
                       middle + i - shift );
          }
          std::move( buffer.begin(),
                     buffer.begin() + batch_size,
                     middle + i - shift );
        }
        std::size_t n_remaining = r_size - i;
        std::move( middle + i, middle + i + n_remaining, buffer.begin() );
        std::size_t shift = 0;
        for ( ; shift <= l_size - r_size; shift += r_size ) {
          std::move( middle + i - shift - r_size,
                     middle + i -shift - r_size + n_remaining,
                     middle + i - shift );
        }
        std::move( buffer.begin(),
                   buffer.begin() + n_remaining,
                   middle + i - shift );
        to     -= shift;
        middle -= shift;
        l_size -= shift;
        continue;
      }

      std::swap_ranges( from, middle, middle );
      break;
    }
    return ( from + r_size );
  }
    
} block_cycle;
\end{lstlisting}
  
  \subsection*{Implementation of the block swap algorithm}
\begin{lstlisting}
constexpr static struct block_swap_t {
    
  template < typename Iterator >
  Iterator operator () ( Iterator from, Iterator middle, Iterator to ) const {
    std::size_t r_size = to - middle;
    std::size_t l_size = middle - from;
    while ( true ) {
      if ( r_size == 0 ) {
        return ( from );
      }
      if ( l_size == 0 ) {
        return ( to );
      }
      if ( r_size >= l_size ) {
        std::swap_ranges( from, middle, middle );
        from = middle;
        middle += l_size;
        r_size -= l_size;
        continue;
      }
      { // 0 < r_size < l_size
        std::swap_ranges( middle, to, middle - r_size );
        to -= r_size;
        middle -= r_size;
        l_size -= r_size;
        continue;
      }
    }
  }
    
} block_swap;
\end{lstlisting}
  
  \subsection*{Implementation of triple reversal and trinity rotation}
  The basic scheme for triple reversal is straightforward.

\begin{lstlisting}
constexpr static struct basic_triple_reverse_t {
public:

  template < typename Iterator >
  Iterator operator () ( Iterator from, Iterator middle, Iterator to ) const {
    std::size_t const r_size = to - middle;
      
    std::reverse( from, middle );
    std::reverse( middle, to );
    std::reverse( from, to );
      
    return ( from + r_size );
  }
    
} basic_triple_reverse;
\end{lstlisting}

  Trinity rotation is more involved.

\begin{lstlisting}
constexpr static struct unchecked_trinity_t {
    
  template < typename A, typename B, typename C >
  void swap3 ( A & a, B & b, C & c ) const {
    auto temp = std::move( a );
    a = std::move( b );
    b = std::move( c );
    c = std::move( temp );
  }

  template < typename A, typename B, typename C, typename D >
  void swap4 ( A & a, B & b, C & c, D & d ) const {
    auto temp = std::move( a );
    a = std::move( b );
    b = std::move( c );
    c = std::move( d );
    d = std::move( temp );
  }
    
  template < typename Iterator >
  Iterator operator () ( Iterator from, Iterator middle, Iterator to ) const {
    std::size_t const l_size = middle - from;
    std::size_t const r_size = to - middle;
      
    Iterator iter_l_beg = from;
    Iterator iter_l_end = middle;
    Iterator iter_r_beg = middle;
    Iterator iter_r_end = to;

    if ( l_size >= r_size ) {
      for ( std::size_t counter = r_size / 2; counter -- > 0; ) {
        -- iter_l_end; -- iter_r_end;
        swap4( *iter_l_end, *iter_l_beg, *iter_r_beg, *iter_r_end );
        ++ iter_l_beg; ++ iter_r_beg;
      }
      for ( std::size_t counter = ( iter_l_end - iter_l_beg ) / 2; counter -- > 0; ) {
        -- iter_l_end; -- iter_r_end;
        swap3( *iter_l_end, *iter_l_beg, *iter_r_end );
        ++ iter_l_beg;
      }
    } else {
      for ( std::size_t counter = l_size / 2; counter -- > 0; ) {
        -- iter_l_end; -- iter_r_end;
        swap4( *iter_l_end, *iter_l_beg, *iter_r_beg, *iter_r_end );
        ++ iter_l_beg; ++ iter_r_beg;
      }
      for ( std::size_t counter = ( iter_r_end - iter_r_beg ) / 2; counter -- > 0; ) {
        -- iter_r_end;
        swap3( *iter_l_beg, *iter_r_beg, *iter_r_end );
        ++ iter_l_beg; ++ iter_r_beg;
      }
    }
    std::reverse( iter_l_beg, iter_r_end );
    return ( from + r_size );
  }

} unchecked_trinity;
\end{lstlisting}
  
  The cases $\Shift=0$ and $\Shift=\Size$ have to be guarded
  against. We employ a generic implementation.

\begin{lstlisting}
constexpr static struct checking_t {
private:
    
  template < typename Rotate >
  struct worker {

    Rotate rotate;

    template < typename Iterator >
    Iterator operator () ( Iterator from, Iterator middle, Iterator to ) const {
      if ( middle == from || middle == to ) {
        return ( from + ( to - middle ) );
      }
      return ( rotate( from, middle, to ) );
    }
      
  };

public:

  template < typename Rotate >
  constexpr worker< Rotate > operator [] ( Rotate rotate ) const {
    return { rotate };
  }
    
} checking;
\end{lstlisting}

  In the case of triple reversal, the cheap check
  avoids a costly operation. Then, we put:
  
\begin{lstlisting}
constexpr static auto trinity = checking[ unchecked_trinity ];
constexpr static auto triple_reverse = checking[ basic_triple_reverse ];
\end{lstlisting}
  
  \subsection*{Buffering}
  Small rotations benefit from using a buffer analogously to
  the early exit strategy mentioned for the block cycle algorithm.
  We can implement that also in a generic fashion:

\begin{lstlisting}
constexpr static struct buffering_t {
private:
    
  template < typename Rotate >
  struct worker {

    Rotate rotate;

    template < typename Iterator >
    Iterator operator () ( Iterator from, Iterator middle, Iterator to ) const {
      std::size_t l_size = middle - from;
      std::size_t r_size = to - middle;
      Iterator result = from + r_size;
        
      if ( middle == from || middle == to ) {
        return ( result );
      }

      constexpr static const unsigned buffer_size = BUFFER_SIZE;
      using value_type = std::decay_t< decltype( *from ) >;
      thread_local std::array< value_type, buffer_size > buffer;
      if ( l_size <= buffer_size ) {
        std::move( from, middle, buffer.begin() );
        std::move( middle, to, from );
        std::move( buffer.begin(), buffer.begin() + l_size, to - l_size );
        return ( result );
      } else if ( r_size <= buffer_size ) {
        std::move( middle, to, buffer.begin() );
        std::move_backward( from, middle, to );
        std::move( buffer.begin(), buffer.begin() + r_size, from );
        return ( result );
      }

      return ( rotate( from, middle, to ) );
    }
      
  };

public:

  template < typename Rotate >
  constexpr worker< Rotate > operator [] ( Rotate rotate ) const {
    return { rotate };
  }
    
} buffering;
\end{lstlisting}

  Then, we put:
  
\begin{lstlisting}
constexpr static auto buffering_triple_reverse = buffering[ basic_triple_reverse ];
constexpr static auto buffering_trinity = buffering[ unchecked_trinity ];
\end{lstlisting}

  For the measurements, we use a buffer size of $256$ items.

  \subsection*{Direct benchmarking}
  We benchmark seven algorithms (block cycling, block
  swapping, buffered and unbuffered trinity rotation, buffered and
  unbuffered triple reverse, and \verb+std::rotate+) with arrays of
  five different element types (\verb+int64_t+, \verb+int32_t+,
  \verb+short+, \verb+char+, and \verb+long double+) all of which are
  trivially movable.

  The measurements were taken on a    Intel(R) Core(TM) i7-10510U CPU @ 1.80GHz
\unskip.
  The cache configuration was as follows:
  \begin{center}
    {\fontsize{9}{12}\selectfont
\begin{verbatim}
NAME ONE-SIZE ALL-SIZE WAYS TYPE        LEVEL SETS PHY-LINE COHERENCY-SIZE
L1d       32K     128K    8 Data            1   64        1             64
L1i       32K     128K    8 Instruction     1   64        1             64
L2       256K       1M    4 Unified         2 1024        1             64
L3         8M       8M   16 Unified         3 8192        1             64
\end{verbatim}}
  \end{center}
  Code was compiled using Clang and {\small GCC} at optimization level~3.
 
  The graphs in Figures~\ref{fig.compare-algs-clang}
  and~\ref{fig.compare-algs-gcc} show runtime in
  nano seconds per byte as a function of the number of bytes in the
  array, i.e., the horizontal axis shows the number of elements times
  the size of the element type. These graphs compare the different
  algorithm.  The graphs in Figures~\ref{fig.compare-types-clang}
  and ~\ref{fig.compare-types-gcc}
  present the same data, but this time we compare how the runtime of a
  fixed algorithm depends on the element type of the sequence.

  For trivially movable types, block cycling outperforms the other
  algorithms for arrays of medium, large, and huge length. For small
  arrays, the results are mixed and depend on the size of the data
  type: for rotating a small array of \verb+char+, block cycling is
  faster than a trinity rotation, whereas for rotating a small array of
  \verb+int64_t+ triple reversal appears to be optimal.
  
  It is notable that, for trivially movable types, the block cycle
  algorithm has a performance almost independent of the element type.
  The effect is most pronounced with Clang, see
  Figure~\ref{fig.compare-types-clang}; but it also shows with {\small
    GCC}. This phenomenon occurs to a slightly lesser degree also in the
  block swapping scheme; but it is not exhibited by the other rotation
  algorithms. It stands to reason that this effect can be attributed to
  two factors: (a) processing batches of adjacent elements and (b)
  preserving the relative order of elements in all move operations. This
  way, a compiler can utilize machine instructions for long words even
  for rotating an array of \verb+char+. If a compiler uses the full
  width of hardware registers in one of the order reversing algorithms,
  some additional code is needed to rearrange the byte order in the
  register.  We also note that \verb+long double+ exhibits somewhat
  anomalous behavior with several algorithms.

  \bibliographystyle{plain}
  \bibliography{rotate}

  \newpage
  \newcommand{\ShowPlot}[3][7cm]{%
    \parbox{#1}{%
      \hbox to #1{\hss #2}

      \vspace*{-4mm}
      
      \null\hfill\null \mbox{#3}\null\hfill\null
    }\hspace{0cm}
  }
  \newenvironment{figpage}{\begin{figure}}{\end{figure}}
  \newcommand{\visualsWithLegend}{%
    \datavisualization
    [
    scientific axes,
    x axis={logarithmic, ticks=few},
    y axis={include value=0, grid},
    visualize as line=blockcycle,
    blockcycle={label in legend={text=block cycle},style={black,mark=*}},
    visualize as line=blockswap,
    blockswap={label in legend={text=block swap},style={black,dashed,mark=*}},
    visualize as line=trinity,
    trinity={label in legend={text=trinity rotate},style={blue,dotted,mark=o}},
    visualize as line=buftrinity,
    buftrinity={label in legend={text=buffering trinity},style={blue,dashed,mark=o}},
    visualize as line=reverse,
    reverse={label in legend={text=triple reverse},style={red,dotted,mark=x}},
    visualize as line=bufreverse,
    bufreverse={label in legend={text=buffering triple reverse},style={red,dashed,mark=x}},
    visualize as line=std,
    std={label in legend={text=std::rotate},style={green,mark=+}},
    ]
  }
  \newcommand{\visuals}{%
    \datavisualization
    [
    scientific axes,
    x axis={logarithmic, ticks=few},
    y axis={include value=0, grid},
    visualize as line=blockcycle,
    blockcycle={style={black,mark=*}},
    visualize as line=blockswap,
    blockswap={style={black,dashed,mark=*}},
    visualize as line=trinity,
    trinity={style={blue,dotted,mark=o}},
    visualize as line=buftrinity,
    buftrinity={style={blue,dashed,mark=o}},
    visualize as line=reverse,
    reverse={style={red,dotted,mark=x}},
    visualize as line=bufreverse,
    bufreverse={style={red,dashed,mark=x}},
    visualize as line=std,
    std={style={green,mark=+}},
    ]
  }
  \begin{figpage}
    \begin{flushleft}
      \ShowPlot{%
      \begin{tikzpicture}[xscale=1.15,yscale=1.35]
        \visuals
        data[set=blockcycle, read from file=clang_timings_block_cycle_long_double.data]
        data[set=blockswap, read from file=clang_timings_block_swap_long_double.data]
        data[set=std, read from file=clang_timings_std_long_double.data]
        data[set=reverse, read from file=clang_timings_triple_reverse_long_double.data]
        data[set=trinity, read from file=clang_timings_trinity_long_double.data]
        data[set=bufreverse, read from file=clang_timings_buffering_triple_reverse_long_double.data]
        data[set=buftrinity, read from file=clang_timings_buffering_trinity_long_double.data]
        ;
      \end{tikzpicture}}{long double}
    \ShowPlot{%
      \begin{tikzpicture}[xscale=1.15,yscale=1.35]
        \visuals
        data[set=blockcycle, read from file=clang_timings_block_cycle_int64.data]
        data[set=blockswap, read from file=clang_timings_block_swap_int64.data]
        data[set=std, read from file=clang_timings_std_int64.data]
        data[set=reverse, read from file=clang_timings_triple_reverse_int64.data]
        data[set=trinity, read from file=clang_timings_trinity_int64.data]
        data[set=bufreverse, read from file=clang_timings_buffering_triple_reverse_int64.data]
        data[set=buftrinity, read from file=clang_timings_buffering_trinity_int64.data]
        ;
      \end{tikzpicture}}{int64}
    \end{flushleft}
    \begin{flushleft}
      \ShowPlot{%
      \begin{tikzpicture}[xscale=1.15,yscale=1.35]
        \visuals
        data[set=blockcycle, read from file=clang_timings_block_cycle_int32.data]
        data[set=blockswap, read from file=clang_timings_block_swap_int32.data]
        data[set=std, read from file=clang_timings_std_int32.data]
        data[set=reverse, read from file=clang_timings_triple_reverse_int32.data]
        data[set=trinity, read from file=clang_timings_trinity_int32.data]
        data[set=bufreverse, read from file=clang_timings_buffering_triple_reverse_int32.data]
        data[set=buftrinity, read from file=clang_timings_buffering_trinity_int32.data]
        ;
      \end{tikzpicture}}{int32}
    \ShowPlot{%
      \begin{tikzpicture}[xscale=1.15,yscale=1.35]
        \visuals
        data[set=blockcycle, read from file=clang_timings_block_cycle_short.data]
        data[set=blockswap, read from file=clang_timings_block_swap_short.data]
        data[set=std, read from file=clang_timings_std_short.data]
        data[set=reverse, read from file=clang_timings_triple_reverse_short.data]
        data[set=trinity, read from file=clang_timings_trinity_short.data]
        data[set=bufreverse, read from file=clang_timings_buffering_triple_reverse_short.data]
        data[set=buftrinity, read from file=clang_timings_buffering_trinity_short.data]
        ;
      \end{tikzpicture}}{short}
    \end{flushleft}
    \begin{flushleft}
      \ShowPlot[12.8cm]{%
      \begin{tikzpicture}[xscale=1.15,yscale=1.35]
        \visualsWithLegend
        data[set=blockcycle, read from file=clang_timings_block_cycle_char.data]
        data[set=blockswap, read from file=clang_timings_block_swap_char.data]
        data[set=std, read from file=clang_timings_std_char.data]
        data[set=reverse, read from file=clang_timings_triple_reverse_char.data]
        data[set=trinity, read from file=clang_timings_trinity_char.data]
        data[set=bufreverse, read from file=clang_timings_buffering_triple_reverse_char.data]
        data[set=buftrinity, read from file=clang_timings_buffering_trinity_char.data]
        ;
      \end{tikzpicture}}{char\hspace*{5cm}\null}
    \end{flushleft}
    \caption{\label{fig.compare-algs-clang}Performance of different rotate algorithms for arrays of
      long double, int64, int32, short,
      and char.
      The $x$-axis shows the size of the array measure in bytes;
      and the $y$-axis shows the runtime in nanoseconds per byte.
      (Code compiled with Clang++ at optimization level 3)}
  \end{figpage}

  \begin{figpage}
    \begin{flushleft}
      \ShowPlot{%
      \begin{tikzpicture}[xscale=1.15,yscale=1.35]
        \visuals
        data[set=blockcycle, read from file=gcc_timings_block_cycle_long_double.data]
        data[set=blockswap, read from file=gcc_timings_block_swap_long_double.data]
        data[set=std, read from file=gcc_timings_std_long_double.data]
        data[set=reverse, read from file=gcc_timings_triple_reverse_long_double.data]
        data[set=trinity, read from file=gcc_timings_trinity_long_double.data]
        data[set=bufreverse, read from file=gcc_timings_buffering_triple_reverse_long_double.data]
        data[set=buftrinity, read from file=gcc_timings_buffering_trinity_long_double.data]
        ;
      \end{tikzpicture}}{long double}
    \ShowPlot{%
      \begin{tikzpicture}[xscale=1.15,yscale=1.35]
        \visuals
        data[set=blockcycle, read from file=gcc_timings_block_cycle_int64.data]
        data[set=blockswap, read from file=gcc_timings_block_swap_int64.data]
        data[set=std, read from file=gcc_timings_std_int64.data]
        data[set=reverse, read from file=gcc_timings_triple_reverse_int64.data]
        data[set=trinity, read from file=gcc_timings_trinity_int64.data]
        data[set=bufreverse, read from file=gcc_timings_buffering_triple_reverse_int64.data]
        data[set=buftrinity, read from file=gcc_timings_buffering_trinity_int64.data]
        ;
      \end{tikzpicture}}{int64}
    \end{flushleft}
    \begin{flushleft}
      \ShowPlot{%
      \begin{tikzpicture}[xscale=1.15,yscale=1.35]
        \visuals
        data[set=blockcycle, read from file=gcc_timings_block_cycle_int32.data]
        data[set=blockswap, read from file=gcc_timings_block_swap_int32.data]
        data[set=std, read from file=gcc_timings_std_int32.data]
        data[set=reverse, read from file=gcc_timings_triple_reverse_int32.data]
        data[set=trinity, read from file=gcc_timings_trinity_int32.data]
        data[set=bufreverse, read from file=gcc_timings_buffering_triple_reverse_int32.data]
        data[set=buftrinity, read from file=gcc_timings_buffering_trinity_int32.data]
        ;
      \end{tikzpicture}}{int32}
    \ShowPlot{%
      \begin{tikzpicture}[xscale=1.15,yscale=1.35]
        \visuals
        data[set=blockcycle, read from file=gcc_timings_block_cycle_short.data]
        data[set=blockswap, read from file=gcc_timings_block_swap_short.data]
        data[set=std, read from file=gcc_timings_std_short.data]
        data[set=reverse, read from file=gcc_timings_triple_reverse_short.data]
        data[set=trinity, read from file=gcc_timings_trinity_short.data]
        data[set=bufreverse, read from file=gcc_timings_buffering_triple_reverse_short.data]
        data[set=buftrinity, read from file=gcc_timings_buffering_trinity_short.data]
        ;
      \end{tikzpicture}}{short}
    \end{flushleft}
    \begin{flushleft}
      \ShowPlot[12.8cm]{%
      \begin{tikzpicture}[xscale=1.15,yscale=1.35]
        \visualsWithLegend
        data[set=blockcycle, read from file=gcc_timings_block_cycle_char.data]
        data[set=blockswap, read from file=gcc_timings_block_swap_char.data]
        data[set=std, read from file=gcc_timings_std_char.data]
        data[set=reverse, read from file=gcc_timings_triple_reverse_char.data]
        data[set=trinity, read from file=gcc_timings_trinity_char.data]
        data[set=bufreverse, read from file=gcc_timings_buffering_triple_reverse_char.data]
        data[set=buftrinity, read from file=gcc_timings_buffering_trinity_char.data]
        ;
      \end{tikzpicture}}{char\hspace*{5cm}\null}
    \end{flushleft}
    \caption{\label{fig.compare-algs-gcc}Performance of different rotate algorithms for arrays of
      long double, int64, int32, short,
      and char.
      The $x$-axis shows the size of the array measure in bytes;
      and the $y$-axis shows the runtime in nanoseconds per byte.
      (Code compiled with GCC at optimization level 3)}
  \end{figpage}

  \newcommand{\altvisualsWithLegend}{
    \datavisualization
    [
    scientific axes,
    y axis={include value=0, grid},
    x axis={logarithmic, ticks=few},
    visualize as line=longd,
    longd={label in legend={text=long double},style={black,dashed,mark=star}},
    visualize as line=longint,
    longint={label in legend={text=int64},style={red,dashed,mark=*}},
    visualize as line=int,
    int={label in legend={text=int32},style={blue,dotted,mark=+}},
    visualize as line=short,
    short={label in legend={text=short},style={black,dashed,mark=x}},
    visualize as line=char,
    char={label in legend={text=char},style={black,dotted,mark=o}},
    ]
  }
  \newcommand{\altvisuals}{
    \datavisualization
    [
    scientific axes,
    y axis={include value=0, grid},
    x axis={logarithmic, ticks=few},
    visualize as line=longd,
    longd={style={black,dashed,mark=star}},
    visualize as line=longint,
    longint={style={red,dashed,mark=*}},
    visualize as line=int,
    int={style={blue,dotted,mark=+}},
    visualize as line=short,
    short={style={black,dashed,mark=x}},
    visualize as line=char,
    char={style={black,dotted,mark=o}},
    ]
  }
  \begin{figpage}
    \begin{flushleft}
      \ShowPlot{%
      \begin{tikzpicture}[xscale=1.18,yscale=1.0]
        \altvisuals
        data[set=longd, read from file=clang_timings_block_cycle_long_double.data]
        data[set=longint, read from file=clang_timings_block_cycle_int64.data]
        data[set=int, read from file=clang_timings_block_cycle_int32.data]
        data[set=short, read from file=clang_timings_block_cycle_short.data]
        data[set=char, read from file=clang_timings_block_cycle_char.data]
        ;
      \end{tikzpicture}}{block cycle}
    \ShowPlot{%
      \begin{tikzpicture}[xscale=1.18,yscale=1.0]
        \altvisuals
        data[set=longd, read from file=clang_timings_block_swap_long_double.data]
        data[set=longint, read from file=clang_timings_block_swap_int64.data]
        data[set=int, read from file=clang_timings_block_swap_int32.data]
        data[set=short, read from file=clang_timings_block_swap_short.data]
        data[set=char, read from file=clang_timings_block_swap_char.data]
        ;
      \end{tikzpicture}}{block swap}
    \end{flushleft}
    \begin{flushleft}
      \ShowPlot{%
      \begin{tikzpicture}[xscale=1.18,yscale=1.0]
        \altvisuals
        data[set=longd, read from file=clang_timings_triple_reverse_long_double.data]
        data[set=longint, read from file=clang_timings_triple_reverse_int64.data]
        data[set=int, read from file=clang_timings_triple_reverse_int32.data]
        data[set=short, read from file=clang_timings_triple_reverse_short.data]
        data[set=char, read from file=clang_timings_triple_reverse_char.data]
        ;
      \end{tikzpicture}}{triple reverse}
    \ShowPlot{%
      \begin{tikzpicture}[xscale=1.18,yscale=1.0]
        \altvisuals
        data[set=longd, read from file=clang_timings_trinity_long_double.data]
        data[set=longint, read from file=clang_timings_trinity_int64.data]
        data[set=int, read from file=clang_timings_trinity_int32.data]
        data[set=short, read from file=clang_timings_trinity_short.data]
        data[set=char, read from file=clang_timings_trinity_char.data]
        ;
      \end{tikzpicture}}{trinity rotate}
    \end{flushleft}
    \begin{flushleft}
      \ShowPlot{%
      \begin{tikzpicture}[xscale=1.18,yscale=1.0]
        \altvisuals
        data[set=longd, read from file=clang_timings_buffering_triple_reverse_long_double.data]
        data[set=longint, read from file=clang_timings_buffering_triple_reverse_int64.data]
        data[set=int, read from file=clang_timings_buffering_triple_reverse_int32.data]
        data[set=short, read from file=clang_timings_buffering_triple_reverse_short.data]
        data[set=char, read from file=clang_timings_buffering_triple_reverse_char.data]
        ;
      \end{tikzpicture}}{triple reverse (buffered)}
    \ShowPlot{%
      \begin{tikzpicture}[xscale=1.18,yscale=1.0]
        \altvisuals
        data[set=longd, read from file=clang_timings_buffering_trinity_long_double.data]
        data[set=longint, read from file=clang_timings_buffering_trinity_int64.data]
        data[set=int, read from file=clang_timings_buffering_trinity_int32.data]
        data[set=short, read from file=clang_timings_buffering_trinity_short.data]
        data[set=char, read from file=clang_timings_buffering_trinity_char.data]
        ;
      \end{tikzpicture}}{trinity rotate (buffered)}
    \end{flushleft}
    \begin{flushleft}\ShowPlot[10.7cm]{%
      \begin{tikzpicture}[xscale=1.18,yscale=1.0]
        \altvisualsWithLegend
        data[set=longd, read from file=clang_timings_std_long_double.data]
        data[set=longint, read from file=clang_timings_std_int64.data]
        data[set=int, read from file=clang_timings_std_int32.data]
        data[set=short, read from file=clang_timings_std_short.data]
        data[set=char, read from file=clang_timings_std_char.data]
        ;
      \end{tikzpicture}}{std::rotate\hspace*{3cm}\null}
    \end{flushleft}
    \caption{\label{fig.compare-types-clang}Performance of different rotate algorithms for arrays.
      The $x$-axis shows the size of the array measure in bytes;
      and the $y$-axis shows the runtime in nanoseconds per byte.
      (Code compiled with Clang at optimization level 3)}
  \end{figpage}

  \begin{figpage}
    \begin{flushleft}
      \ShowPlot{%
      \begin{tikzpicture}[xscale=1.18,yscale=1.0]
        \altvisuals
        data[set=longd, read from file=gcc_timings_block_cycle_long_double.data]
        data[set=longint, read from file=gcc_timings_block_cycle_int64.data]
        data[set=int, read from file=gcc_timings_block_cycle_int32.data]
        data[set=short, read from file=gcc_timings_block_cycle_short.data]
        data[set=char, read from file=gcc_timings_block_cycle_char.data]
        ;
      \end{tikzpicture}}{block cycle}
    \ShowPlot{%
      \begin{tikzpicture}[xscale=1.18,yscale=1.0]
        \altvisuals
        data[set=longd, read from file=gcc_timings_block_swap_long_double.data]
        data[set=longint, read from file=gcc_timings_block_swap_int64.data]
        data[set=int, read from file=gcc_timings_block_swap_int32.data]
        data[set=short, read from file=gcc_timings_block_swap_short.data]
        data[set=char, read from file=gcc_timings_block_swap_char.data]
        ;
      \end{tikzpicture}}{block swap}
    \end{flushleft}
    \begin{flushleft}
      \ShowPlot{%
      \begin{tikzpicture}[xscale=1.18,yscale=1.0]
        \altvisuals
        data[set=longd, read from file=gcc_timings_triple_reverse_long_double.data]
        data[set=longint, read from file=gcc_timings_triple_reverse_int64.data]
        data[set=int, read from file=gcc_timings_triple_reverse_int32.data]
        data[set=short, read from file=gcc_timings_triple_reverse_short.data]
        data[set=char, read from file=gcc_timings_triple_reverse_char.data]
        ;
      \end{tikzpicture}}{triple reverse}
    \ShowPlot{%
      \begin{tikzpicture}[xscale=1.18,yscale=1.0]
        \altvisuals
        data[set=longd, read from file=gcc_timings_trinity_long_double.data]
        data[set=longint, read from file=gcc_timings_trinity_int64.data]
        data[set=int, read from file=gcc_timings_trinity_int32.data]
        data[set=short, read from file=gcc_timings_trinity_short.data]
        data[set=char, read from file=gcc_timings_trinity_char.data]
        ;
      \end{tikzpicture}}{trinity rotate}
    \end{flushleft}
    \begin{flushleft}
      \ShowPlot{%
      \begin{tikzpicture}[xscale=1.18,yscale=1.0]
        \altvisuals
        data[set=longd, read from file=gcc_timings_buffering_triple_reverse_long_double.data]
        data[set=longint, read from file=gcc_timings_buffering_triple_reverse_int64.data]
        data[set=int, read from file=gcc_timings_buffering_triple_reverse_int32.data]
        data[set=short, read from file=gcc_timings_buffering_triple_reverse_short.data]
        data[set=char, read from file=gcc_timings_buffering_triple_reverse_char.data]
        ;
      \end{tikzpicture}}{triple reverse (buffered)}
    \ShowPlot{%
      \begin{tikzpicture}[xscale=1.18,yscale=1.0]
        \altvisuals
        data[set=longd, read from file=gcc_timings_buffering_trinity_long_double.data]
        data[set=longint, read from file=gcc_timings_buffering_trinity_int64.data]
        data[set=int, read from file=gcc_timings_buffering_trinity_int32.data]
        data[set=short, read from file=gcc_timings_buffering_trinity_short.data]
        data[set=char, read from file=gcc_timings_buffering_trinity_char.data]
        ;
      \end{tikzpicture}}{trinity rotate (buffered)}
    \end{flushleft}
    \begin{flushleft}\ShowPlot[10.7cm]{%
      \begin{tikzpicture}[xscale=1.18,yscale=1.0]
        \altvisualsWithLegend
        data[set=longd, read from file=gcc_timings_std_long_double.data]
        data[set=longint, read from file=gcc_timings_std_int64.data]
        data[set=int, read from file=gcc_timings_std_int32.data]
        data[set=short, read from file=gcc_timings_std_short.data]
        data[set=char, read from file=gcc_timings_std_char.data]
        ;
      \end{tikzpicture}}{std::rotate\hspace*{3cm}\null}
    \end{flushleft}
    \caption{\label{fig.compare-types-gcc}Performance of different rotate algorithms for arrays.
      The $x$-axis shows the size of the array measure in bytes;
      and the $y$-axis shows the runtime in nanoseconds per byte.
      (Code compiled with GCC at optimization level 3)}
  \end{figpage}
\end{document}